\documentclass{article}
\usepackage{arxiv}
\usepackage{amsthm} 
\usepackage{amssymb}
\usepackage{amsmath}
\usepackage{empheq}
\usepackage{amscd}
\usepackage{graphicx}
\usepackage{graphics}
\usepackage[noadjust]{cite}
\usepackage{amsthm}
\usepackage{caption}
\usepackage{multirow}
\usepackage{array}
\usepackage{rotating}
\usepackage[norelsize,ruled]{algorithm2e}
\usepackage{hyperref}

\usepackage{indentfirst}
\usepackage{tikz}
\usepackage{calc}
\usepackage{epsfig}
\usepackage[numbers]{natbib}
\usepackage[makeroom]{cancel}
\usepackage{multicol}
\usepackage{csquotes}
\usepackage{enumitem}
\usepackage{nicefrac}
\usepackage{hyperref}       
\hypersetup{colorlinks,linkcolor={green},citecolor={green},urlcolor={black}}
\usepackage{optidef}

\usepackage{nicefrac}

\usepackage{soul}

\newcommand{\eref}[1]{equation \ref{#1}}                  

\newcommand{\fref}[1]{Figure \ref{#1}}                  
\DeclareMathOperator*{\minimize}{minimize}
\DeclareMathOperator*{\maximize}{maximize}

\newcommand{\bd}{\bold{d}}

\newcommand{\bm}{\bold{m}}

\newcommand{\bb}{\bold{b}}
\newcommand{\bP}{\bold{P}}
\newcommand{\bA}{\bold{A}}
\newcommand{\bu}{\bold{u}}

\begin{document}
\title{Direct inversion of data space Hessian for efficient time-domain extended-source waveform inversion using the multiplier method}

\author{  \href{https://orcid.org/0000-0003-3372-1800}{\hspace{1mm}Mahdi Sonbolestan}\\
  Institute of Geophysics, Polish Academy of Sciences, Warsaw, Poland\\
  \texttt{msonbolestan@igf.edu.pl} \\
  \AND
\href{https://orcid.org/0000-0002-9879-2944}{\hspace{1mm}Ali Gholami} \\
  Institute of Geophysics, Polish Academy of Sciences, Warsaw, Poland\\
  \texttt{agholami@igf.edu.pl} \\ 
  }

\renewcommand{\shorttitle}{Direct data space Hessian inversion ~~~~~~~~~~~~~ Sonbolestan and Gholami}

\maketitle

\begin{abstract}
The augmented Lagrangian (AL) method has been successfully applied for solving the full waveform inversion (FWI) problem. In AL-based FWI, the Lagrange multipliers serve as source extensions, offering several advantages to the inversion, such as improved robustness to cycle skipping, faster convergence, and simplified penalty parameter tuning. Time-domain applications of this method have been enabled by reformulating the optimization problem in the data space, significantly reducing memory requirements by projecting source-side multipliers into the data space. These data-side multipliers act as data extensions, effectively expanding the data space. A key challenge in these methods lies in computing the data-side multipliers, which involves solving a linear system to deblur the data residuals using the data-space Hessian matrix before it serves as the adjoint source. This Hessian matrix is prohibitively large to construct and invert explicitly. Iterative Krylov methods can be applied to solve this system as inner iterations, but they require two PDE solves per inner iteration per source, leading to significant computational costs. In this work, we present a key improvement to extended waveform inversion based on multiplier methods. We propose a novel approach that significantly reduces the computational cost of Hessian inversion. The method computes receiver-side Green functions in the time domain and directly constructs frequency-domain Hessian matrices for all required frequencies. These Hessian matrices, with dimensions equal to the number of receivers, can be computed, inverted, and stored in memory. Once constructed, they can be used simultaneously for all sources, further enhancing efficiency. Numerical experiments on benchmark models demonstrate the substantial computational gains achieved by the proposed method, highlighting its effectiveness and practicality for extended-source FWI in the time domain.
\end{abstract}
\section{Introduction}
\noindent 
Full waveform inversion (FWI) is an advanced inversion technique for high-resolution imaging of subsurface properties, such as velocity, density, and anisotropy, using surface measurements of the seismic waveﬁeld. It iteratively minimizes the misfit between observed seismic data and synthetic data generated through wave-equation modeling. Since its introduction in the 1980s through the pioneering work of Lailly \citep{Lailly_1983_SIP} and Tarantola \citep{Tarantola_1984_ISR}, who applied gradient descent methods to update model parameters by minimizing discrepancies between observed and simulated data, FWI has undergone significant advancements. In the 1990s, the frequency-domain formulation proposed by \citet{Pratt_1998_GNF} improved computational efficiency for multiscale, multisource inversion settings. However, a key limitation of frequency-domain implementation is the difficulty of solving large-scale monochromatic wavefield problems due to ill-conditioning of the Helmholtz operator \citep{Ernst_2011_WDS}.
To address this, \citet{Bunks_1995_MSW} introduced the multiscale inversion approach in the time domain, employing sequential applications of low-pass ﬁlters and/or time windowing techniques to prioritize inversion of early arrivals. This strategy enables FWI to be applied effectively to large-scale problems using time-domain finite-difference solvers, which ensure scalability. Although multiscale inversion improves robustness, FWI remains highly sensitive to initial models, necessitating the development of more robust algorithms capable of converging from inaccurate initial models.
Ensuring the convergence of FWI algorithms is challenging due to cycle skipping, which occurs when the initial model fails to predict the observed wavefield within half a cycle. In such cases, the iterative algorithm may converge to a geologically meaningless local minimum of the misfit function \citep{Metivier_2017_TRU}. In recent years, considerable attention has been devoted to developing advanced inversion techniques aimed at mitigating cycle skipping and enhancing the reliability of FWI
 \citep[e.g.,][]{Warner_2016_AWI,Symes_2020_WRI,Rizzuti_2021_ADF,Gholami_2022_EFW,Metivier_2022_RES,
Barnier_2023_FWIp,Barnier_2023_FWIt,Symes_2024_NIM}. 

Extended-source FWI methods have proven to be robust approaches for accurately estimating subsurface parameters. The fundamental idea is to impose the wave-equation constraint as a soft constraint via a penalty term in the objective function. Unlike the classical reduced-form FWI, where this constraint must be satisfied from the initial iteration, extended inversion only requires it to be satisfied at convergence \citep{Abubakar_2009_FDC,VanLeeuwen_2013_MLM,Huang_2018_VSE,Aghamiry_2019_IWR,Symes_2020_FWI}. This relaxation provides greater flexibility in fitting the data from any initial model regardless of its accuracy, thereby mitigating cycle skipping \citep{Operto_2023_FWI}. In this formulation, model space errors are effectively projected onto the source space. The added degree of freedom in the source term improves the conditioning of the inverse problem. However, this extended approach introduces additional complexities, including the challenge of managing the larger solution space and dealing with approximate physics, which necessitates solving an augmented wave equation or computing the source extension term \citep{Aghamiry_2020_AED}.
Ensuring a gradual reduction in wave-equation violations while progressively enforcing physical constraints is another key challenge. This issue is efficiently addressed by the multiplier method, which is specifically designed for constrained optimization problems \citep{Aghamiry_2019_IWR,Hestenes_1969_MAG,Powell_1969_NLC}. By incorporating Lagrange multipliers into the augmented Lagrangian objective function, this method provides a flexible FWI framework that combines the robustness of extended inversion with an efficient mechanism for enforcing constraints.

Extended inversion is more easily implemented in the frequency domain due to the advantages of matrix algebra methods. However, its time-domain implementation has also been investigated \citep{Rizzuti_2021_ADF,Gholami_2022_EFW,Guo_2024_TDE,Lin_2023_FWR,Symes_2020_WRI}.
In these methods, the model update is also computed through the cross-correlation of forward and backward wavefields, similar to reduced FWI but involving different wavefields. Unlike reduced FWI—where the forward wavefield results from the physical source and the backward wavefield is obtained by backpropagating the data residuals as the adjoint source—in extended FWI, the forward and backward wavefields are coupled. They arise as the solution to a coupled two-by-two block linear system, known as a saddle point system, which represents a regularized extension of the system used in the reduced FWI. Two main strategies can be adopted to solve this system: the wavefield-oriented approach and the multiplier-oriented approach, each emphasizing either wavefield reconstruction or multiplier estimation as the central objective, offering flexibility in algorithm design \citep[see][ for more details]{Gholami_2024_FWI}.
Time-domain implementations of extended FWI methods, using the multiplier-oriented approach, rely heavily on the precise estimation of Lagrange multipliers, which are used to construct both the forward and backward wavefields. A major computational challenge in this process is solving a linear system to deblur the data residuals using the data-space Hessian matrix before it serves as the adjoint source \citep{Gholami_2022_EFW}. This Hessian matrix also functions as a data weighting matrix, linking extended FWI to reduced FWI through a weighted data misfit norm.

Let $\delta\bold{d}_s$ denote the data residual, defined as the difference between the observed and predicted data for a given source $s$, obtained by reshaping an $N_t \times N_r$ matrix of residual traces into a column vector (where $N_t$ and $N_r$ represent the number of time samples and receivers, respectively). In extended FWI, the adjoint source $\delta\bold{d}_s^e$ is related to the standard data residual $\delta\bold{d}_s$ through the linear system $\bold{Q}\delta\bold{d}_s^e = \delta\bold{d}_s$, where $\bold{Q}$, of size $N_t N_r \times N_t N_r$, is the data-space Hessian matrix \citep[see][ their equation 35]{Gholami_2022_EFW}.
Constructing this Hessian matrix requires $N_r$ PDE solves. However, its size makes explicit construction and inversion computationally prohibitive.
To address this challenge, \citet{Gholami_2022_EFW} approximated the Hessian using a scaled identity matrix, while \citet{Lin_2023_FWR} discussed various theoretical formulations of the covariance matrix. \citet{Guo_2024_TDE} proposed an approximation using matching filters.
These approximations can also serve as preconditioners to accelerate Krylov subspace methods—a class of iterative algorithms used for solving large linear systems through matrix-vector multiplications \citep{Nocedal_2006_NO}. Compared to direct methods, iterative methods require less memory, making them well-suited for large-scale problems. Common examples include the Conjugate Gradient (CG) algorithm and Generalized Minimal Residual Method (GMRES), both of which are frequently employed in practice.
However, iterative methods still require $2 N_{\text{cg}} N_s$ PDE solves, where $N_{\text{cg}}$ is the number of CG iterations and $N_s$ is the number of sources, leading to a significant computational burden \citep{Aghamiry_2020_AED,Sonbolestan_2022_ORD}.

In this work, we propose an efficient direct inversion approach that significantly reduces the computational cost of implementing the data-space Hessian by leveraging the computational advantages of time-domain finite-difference solvers and the separability of frequency-domain inversion. In the proposed hybrid-domain method, Green functions are simulated in the time domain, while the Hessian inversion is performed in the frequency domain.
The hybrid-domain approach was initially proposed by \citep{Sirgue_2008_FDW} in the context of frequency-domain reduced-space FWI, where the wavefield simulations were conducted using time-domain finite-difference methods, followed by gradient calculations in the frequency domain. In contrast, in this work, we implemented a time-domain FWI where the simulation of Green functions is performed efficiently while the Hessian inversion is performed in the frequency domain. This approach replaces the inversion of a large matrix in the time domain with the inversion of a set of smaller matrices in the frequency domain, enhancing efficiency.  The data-space Hessian matrix is structured as a block matrix with $N_r \times N_r$ blocks, each of size $N_t \times N_t$, effectively capturing correlations between data residuals in both time and space. Each block is computed by cross-correlating receiver-side Green functions associated with pairs of receivers. As a result, constructing the full Hessian requires only $N_r$ PDE solves. The convolutional structure of these blocks enables efficient computation of the inverse of the Hessian using the Fast Fourier Transform (FFT). Furthermore, while these blocks are not perfectly diagonalizable in the Fourier domain due to limited time recordings, the elements on the main diagonal of each block provide a sufficiently accurate approximation, allowing efficient storage by retaining only these components. This stored frequency information facilitates the direct computation of frequency-domain Hessian matrices of size $N_r \times N_r$ for all required frequencies, eliminating the need to construct the full Hessian matrix in the time domain. Consequently, the $N_r \times N_r$ Fourier matrices can be directly inverted, bypassing the need for iterative methods.


\section{Theory}
For the analysis in this paper, we assume wave propagation in acoustic media with variable squared slowness $\bold{m}(\bold{x})$ and constant density and isotropic point sources.
The seismic wavefield $\bold{u}_s(t,\bold{x})$ due to a point source at $\bold{x}_s$ satisfies the wave equation
\begin{equation} \label{wave_eq}
\bold{m}(\bold{x})\frac{\partial^2}{\partial t^2}\bold{u}_s(t,\bold{x}) - {\nabla}^2 \bold{u}_s(t,\bold{x})=\delta(\bold{x-x}_s)f(t),
\end{equation}
where $t$ and $\bold{x}$ are respectively the time and space coordinates, ${\nabla}^2$ denotes the Laplacian, $\delta$ is the delta function, $\bold{x}_s$ is the source location, and $f(t)$ is the source signature.
Solving equation \ref{wave_eq} for the wavefield and then sampling it at the receiver locations, by the sampling operator $\bold{P}$, gives the recorded data $\bold{d}_s=\bold{P}\bold{u}_s=\bold{u}_s(\bold{x}_r)$, where $\bold{x}_r$ denote the receiver location.

Taking into account data uncertainty, FWI can be formulated as a nonlinearly constrained optimization problem that estimates the model parameters $\bold{m}$ from locally recorded data $\bold{d}_s, s=1, \dots, N_s$. This is achieved by minimizing the data misfit, defined as the squared Euclidean norm of the data residuals, while ensuring that the wave equation is satisfied for all sources:
\begin{mini} 
{\bold{m},\{\bold{u}_{s}\}_{s=1}^{{N}_s}}{\frac12\sum_{s=1}^{N_s} \| \bold{d}_s-\bold{P}\bold{u}_s \|_2^2}
{\label{main}}{}
\addConstraint {\bold{A(m)u}_{s}}{=\bold{b}_s, s=1,...,N_s. }
\end{mini}
In this equation, $\bold{A(m)}$ is the wave equation operator constructed with sufficient accuracy and appropriate boundary conditions, and $\bold{b}_s$ are the source terms.

The Lagrangian multiplier method provides a standard framework for solving the constrained problem \eqref{main}.  It is based on the optimization of the augmented Lagrangian function as 
\begin{equation} \label{minmax}
\minimize_{\bold{m},\{\bold{u}_{s}\}_{s=1}^{N_s}} \maximize_{\{\boldsymbol{\nu}_{s}\}_{s=1}^{N_s}}  \sum_{s=1}^{N_s} \mathcal{L}(\bold{m, u}_{s}, \boldsymbol{\nu}_{s};\bold{b}_s,\bold{d}_s),
\end{equation}  
where
\begin{equation} \label{AL}
\mathcal{L}(\bold{m, u}, \boldsymbol{\nu};\bold{b},\bold{d}) = 
\frac12\|\bold{d}-\bold{P}\bold{u}\|_2^2
- \langle \boldsymbol{\nu}, \bold{A(m)u} - \bold{b} \rangle 
+ \frac{\mu}{2} \|\bold{A(m)u} - \bold{b}\|_2^2.
\end{equation} 
where $\boldsymbol{\nu}_s$ are the source-side Lagrange multipliers, $\langle\cdot,\cdot \rangle$ denotes the inner product, and $\mu>0$ is the penalty parameter.
The pseudocode for solving \eref{minmax} is presented in Algorithm \ref{alg:AL_time} \citep[see][ for the details]{Gholami_2022_EFW}. 

In Algorithm \ref{alg:AL_time}, the standard data residual is defined as $\delta \bold{d}_s=\bold{d}_s-\bold{S}\bold{b}_s$, where $\bold{S}=\bold{P}\bold{A(m)}^{-1}$ is the forward modeling operator.
The source extension $\delta \bb_s$ is obtained as the damped least-squares solution to the underdetermined system $\bold{S}\delta \bb_s=\delta \bold{d}_s$. This solution is given by $\delta \bb_s=\bold{S}^T\delta \bold{d}_s^e$, where the superscript $T$ denotes matrix transposition and adjoint source $\delta \bold{d}_s^e$ is defined as
\begin{equation} \label{Qsystem}
\delta \bold{d}^e_s = \bold{Q}^{-1} \delta \bold{d}_s,
\end{equation}
where $\bold{Q}$ is the so-called data-space Hessian matrix, defined as
\begin{equation} \label{Qmatrix}
\bold{Q}=\bold{S} \bold{S}^T + \mu \bold{I}.
\end{equation} 
Here, the penalty parameter $\mu>0$ acts as a regularization term to stabilize the matrix inversion. The source extension 
$\delta \bb_s$, when added to the physical source, enables accurate prediction of the data through the forward wavefield $\bold{u}_s^e$, regardless of the velocity model's accuracy. The goal of the inversion is to iteratively reduce the energy of the source extension by improving the accuracy of the velocity model.
A central computational bottleneck in this approach lies in solving \eref{Qsystem} to compute $\delta \bold{d}^e_s$.
In addition to constructing the source extension, the computed $\delta \bold{d}^e_s$ are also used to update the data-side Lagrange multipliers as  $ \Delta \bd_s = \Delta \bd_s + \delta \bd^e_s $, and to compute the backpropagating wavefields $ \boldsymbol{\lambda}_s = \bold{S}^T [\Delta \bd_s + \delta \bd^e_s] $. Finally, $\text{proj}_\mathcal{C}(\bm)$ denotes the projection operator which projects the model onto the desired region defined by bounding constraints.

\begin{algorithm}[hbt!]
\caption{Multiplier-based time-domain extended FWI  using data-space Hessian} \label{alg:AL_time}
\textbf{Input:} $\bd$ (data), $\bm_0$ (initial model parameters), $\bb$ (source wavelet).  \\
\textbf{Initialize:} Set: $k = 0, \ \Delta \bd = 0.$ \\
\While{convergence criteria not satisfied}{
Calculate Hessian matrices $\hat{\bold{Q}}({\omega})^{-1}$ for all required frequencies using Algorithm \ref{alg:Hessian}\\
\For{all s}{
$\delta \bd_s \gets  \bd_s-\bP\bA(\bm_k)^{-1} \bb_s$ \\
compute $\delta \bd^e_s = \bold{Q}^{-1}\delta \bd_s $ in the frequency domain\\
$\delta \bb_s \gets  \bA(\bm_k)^{-T}\bP^T \delta \bd_s^e $ \\
$\bu^e_s \gets \bA(\bm_k)^{-1}(\bb_s + \delta \bb_s)$ \\
$\lambda_s \gets \bA(\bm_k)^{-T}\bP^T (\Delta \bd_s + \delta \bd_s^e) $ \\
$\Delta \bd_s \gets \Delta \bd_s + \delta \bd^e_s $  
}
$\bm_{k+1} \gets \bm_{k} -  \frac{\sum_{s=1}^{N_s}\langle \boldsymbol{\lambda}_s , \ \partial_{tt}\bu^e_s \rangle}{\sum_{s=1}^{N_s}\langle \partial_{tt}\bu^e_s , \ \partial_{tt}\bu^e_s \rangle}$  \\
$\bm_{k+1} \gets \text{proj}_\mathcal{C}(\bm_{k+1}) $\\
$ k \gets k+1 $
}
\end{algorithm}

With an efficient Hessian implementation, the algorithm achieves both memory and computational efficiency. In the following subsection, we analyze the detailed structure of the Hessian and develop a hybrid-domain approach that enables its direct inversion. For simplicity of equations, we ignore the index $s$ for the source, but the algorithm can be easily generalized for more than one source.

\subsection{Direct Hessian-inverse computation} 
For a common source gather $\bold{D}=[\bold{d}^1\vert \bold{d}^2\vert\cdots\vert \bold{d}^{N_r}]$ of size $N_t\times N_r$, there are two common conventions for reshaping it into a vector.
The first is to stack the columns of $\bold{D}$ into a single column vector:
\begin{equation}
\bold{d}=\text{vec}(\bold{D})=
 \begin{pmatrix}
     \bold{d}^1\\
     \bold{d}^2\\
     \vdots\\
     \bold{d}^{N_r}\\
 \end{pmatrix},
\end{equation}
where $\text{vec}(\cdot)$ denotes the vectorization operator that stacks the columns of a matrix into a column vector.
In the second convention, we apply the operator $\text{vec}(\cdot)$ to the transpose of the matrix, $\bold{D}^T$, $\text{vec}(\bold{D}^T)$. These two conventions are related by a permutation matrix $\bold{\Pi}$, which transforms $\text{vec}(\bold{D})$ into $\text{vec}(\bold{D}^T)$. 

The formulation of this paper is based on the first convention; however, we may switch between these two whenever required.
For the first convention, the sampling matrix $\bold{P}$ samples the wavefield for all of the times of one receiver and then the next receiver after that. The resulting sampling matrix in this convention will consist of $N_r$ blocks vertically stacked, in which each block is a $N_t \times N_t N_x$ matrix (where $N_x$ represents the number of model grid points) responsible for sampling all the times for one receiver, one after the other in the appropriate receiver locations. Below you can see the structure of the sampling matrix $\bold{P}$:
\begin{equation}
\bold{P} =
\begin{pmatrix}
\bold{P}^{1} \\
\bold{P}^2  \\
\vdots \\
\bold{P}^{N_r} \\
\end{pmatrix}.
\end{equation}
Each row of block $\bold{P}^j$ represents a delta function associated to the time $t_n$, for $n=1,...,N_t$, and receiver position $\bold{x}_r^j$, symbolized as $\delta(t-t_{n},\bold{x-x}_r^j)$. 

The block structure of the sampling operator induces a corresponding block structure in the Hessian matrix.
Ignoring the regularization term in \eref{Qmatrix} for simplicity, each block $(i,j)$, where $i,j=1,...,N_r$, of the Hessian matrix is given by:
\begin{equation} \label{Hij}
    \bold{Q}^{i,j} = \bold{P}^{i} \bold{A}^{-1}({\bold{P}^{j}}\bold{A}^{-1})^T = \sum_{\bold{x}} \bold{T}(\bold{g}^i(:,\bold{x}))\bold{T}(\bold{g}^j(:,\bold{x}))^T,
\end{equation}
where $\bold{A} \equiv \bold{A}(\bm) $, $\bold{T}(\bold{v})$ is a lower-triangular Toeplitz matrix with vector $\bold{v}$ of size $N_t$ as its first column, and $\bold{g}^j(t,\bold{x})$ is the Green function associated with the receiver location $\bold{x}^j_r$; that is, the wavefield generated by a Dirac delta source placed at the receiver location. Here,  $t$ denotes time and $\bold{x}$ is a point in the medium.
We see that each block of the Hessian matrix is the sum of the products of two Toeplitz matrices over space. The direct multiplication of Toeplitz matrices requires $O(N_t^3)$ operations. 
Consequently, constructing a single block involves $O(N_x N_t^3)$ operations, and building the full Hessian matrix requires $O(N_x N_r^2 N_t^3)$ operations---assuming the Green functions have already been computed, which itself entails solving $N_r$ wave equations. Therefore, direct computation of the Hessian using \eref{Hij} is impractical.

To improve computational efficiency, we replace each Toeplitz matrix with a larger circulant matrix, which can be efficiently manipulated using FFT techniques without losing any information.
To achieve this, we embed $\bold{Q}^{i,j}$ in \eqref{Hij} into a larger system of the form:
\begin{align}\label{Hij_embed}
\bold{Q}_{\text{circ}}^{i,j} 
&= \sum_{\bold{x}}
\overbrace{
   \begin{pmatrix}
   \bold{T}(\bold{g}^i(:,\bold{x})) & \bold{B}(\bold{g}^i(:,\bold{x}))\\
     \bold{B}(\bold{g}^i(:,\bold{x})) & \bold{T}(\bold{g}^i(:,\bold{x}))
\end{pmatrix}
}^{\bold{C}\left(\bold{T}(\bold{g}^i(:,\bold{x}))\right)}
\begin{pmatrix}
   \bold{I} & \bold{0}\\
     \bold{0} & \bold{0}
\end{pmatrix}
\overbrace{
  \begin{pmatrix}
   \bold{T}(\bold{g}^j(:,\bold{x})) & \bold{B}(\bold{g}^j(:,\bold{x}))\\
     \bold{B}(\bold{g}^j(:,\bold{x})) & \bold{T}(\bold{g}^j(:,\bold{x}))
\end{pmatrix}^T
}^{\bold{C}\left(\bold{T}(\bold{g}^j(:,\bold{x}))\right)^T}\\
&= \sum_{\bold{x}}
   \begin{pmatrix}
   \bold{T}(\bold{g}^i(:,\bold{x}))\bold{T}(\bold{g}^j(:,\bold{x}))^T & \bold{T}(\bold{g}^i(:,\bold{x}))\bold{B}(\bold{g}^j(:,\bold{x}))^T\\
    \bold{B}(\bold{g}^i(:,\bold{x}))\bold{T}(\bold{g}^j(:,\bold{x}))^T& \bold{B}(\bold{g}^i(:,\bold{x}))\bold{B}(\bold{g}^j(:,\bold{x}))^T
\end{pmatrix},\label{2by2}
\end{align}
where matrix $\bold{B}(\bold{g}^i(:,\bold{x}))$ is chosen to ensure that $\bold{C}\left(\bold{T}(\bold{g}^i(:,\bold{x}))\right)$ is a circulant matrix. From \eref{2by2}, we observe that $\bold{Q}^{i,j}_{\text{circ}}$ is a $2 \times 2$ block matrix, where its top-left block corresponds to the desired matrix $\bold{Q}^{i,j}$.

Circulant matrices are diagonalizable by the Fourier matrix \citep{Gray_2006_TCM}. Specifically, let $\bold{F}$ be the Fourier matrix of size $2N_t \times 2N_t$, then the circulant matrix $\bold{C}\left(\bold{T}(\bold{g}^i(:,\bold{x}))\right)$ can be decomposed as:  
\begin{equation}
\bold{C}\left(\bold{T}(\bold{g}^i(:,\bold{x}))\right) = \bold{F}^T \text{diag}(\hat{\bold{g}}^i(:,\bold{x})) \bold{F},
\end{equation}
where $\hat{\bold{g}}^i(\omega,\bold{x})$ is the $2N_t$-length Fourier transform of $\bold{g}^i(t,\bold{x})$.  
This property enables a convolution-based representation of the matrix, leading to the following 2D convolution form:  
\begin{align} 
\bold{Q}_{\text{circ}}^{i,j} 
&=\sum_{\bold{x}} \bold{F}^T \text{diag}(\hat{\bold{g}}^i(:,\bold{x})) \bold{F}
\begin{pmatrix}
   \bold{I} & \bold{0}\\
     \bold{0} & \bold{0}
\end{pmatrix}
\bold{F}^T \text{diag}(\hat{\bold{g}}^j(:,\bold{x}))^T \bold{F}\nonumber\\
&=\sum_{\bold{x}} \bold{F}^T \text{diag}(\hat{\bold{g}}^i(:,\bold{x})) 
   \hat{\bold{I}}  \text{diag}(\hat{\bold{g}}^j(:,\bold{x}))^T \bold{F}\nonumber\\
&=\sum_{\bold{x}} \bold{F}^T \left( 
   \hat{\bold{I}} \circ \left[\hat{\bold{g}}^i(:,\bold{x}) \hat{\bold{g}}^j(:,\bold{x})^T\right]\right) \bold{F} \nonumber\\
&= \bold{F}^T \left( 
   \hat{\bold{I}} \circ \left[\sum_{\bold{x}}\hat{\bold{g}}^i(:,\bold{x}) \hat{\bold{g}}^j(:,\bold{x})^T\right]\right) \bold{F} \nonumber\\
   & = \bold{I} \ast \  \bold{G}^{i,j}.
\end{align}
In this equation, $\hat{\bold{I}}=\bold{F} \left( \begin{smallmatrix} \bold{I} & \bold{0} \\ \bold{0} & \bold{0} \end{smallmatrix} \right) \bold{F}^T$, $\circ$ denotes an element-wise product, $\ast $ denotes 2D convolution with zero-padding and $\bold{G}^{i,j} = \sum_{\bold{x}} \bold{g}^i(:,\bold{x}) \bold{g}^j(:,\bold{x})^T$ is the outer product of two Green functions associated with receivers $i$ and $j$. Summing the elements of $\bold{G}^{i,j}$ along its descending diagonals yields the cross-correlation between the two Green functions.
Note that both the identity matrix $\bold{I}$ and the $\bold{G}^{i,j}$ are $N_t \times N_t$ matrices. However, their convolution produces a $(2N_t - 1) \times (2N_t - 1)$ matrix. The top-left $N_t \times N_t$ portion of this resulting matrix corresponds to the desired matrix $\bold{Q}^{i,j}$. The 2D convolution with an identity matrix and cropping the top left $N_t \times N_t$ can be interpreted as summing up the diagonals of the matrix as
\begin{equation} \label{Q}
    \bold{Q}^{i,j}=
    \begin{bmatrix}
    \bold{G}^{i,j}_{1,1} & \bold{G}^{i,j}_{1,2}& \bold{G}^{i,j}_{1,3}& \cdots & \bold{G}^{i,j}_{1,N_t} \\
    \bold{G}^{i,j}_{2,1} & {\displaystyle\sum_{n=1}^2\bold{G}^{i,j}_{n,n}}& {\displaystyle\sum_{n=1}^2\bold{G}^{i,j}_{n,n+1}}& 
    \cdots & 
    {\displaystyle\sum_{n=1}^2 \bold{G}^{i,j}_{n,n+N_t-2}} \\
    \bold{G}^{i,j}_{3,1} &
    {\displaystyle\sum_{n=1}^2\bold{G}^{i,j}_{n+1,n}}&
    {\displaystyle\sum_{n=1}^3\bold{G}^{i,j}_{n,n}}&
    \ddots &
    \vdots\\
    \vdots & \vdots & \ddots & \ddots & 
    { \displaystyle\sum_{n=1}^{N_t-1} \bold{G}^{i,j}_{n,n+1}} \\
    \bold{G}^{i,j}_{N_t,1} &
    {\displaystyle\sum_{n=1}^2\bold{G}^{i,j}_{n+N_t-2,n}} &
    \cdots &
    {\displaystyle\sum_{n=1}^{N_t-1}\bold{G}^{i,j}_{n+1,n}}&
    {\displaystyle\sum_{n=1}^{N_t}\bold{G}^{i,j}_{n,n}}&
    \end{bmatrix}.
\end{equation}
This special structure of the matrix $\bold{Q}^{i,j}$ arises due to the finite recording time. As $t \to \infty$, the amplitude of the Green functions diminishes, and $\bold{Q}^{i,j}$ converges to a Toeplitz matrix composed of $2N_t - 1$ unique values, corresponding to the cross-correlation of the Green functions summed over space.

\subsubsection{Fourier domain Hessian}
Although a Toeplitz matrix is not exactly diagonal in the Fourier domain, it can be closely approximated by a circulant matrix, which is diagonal in the Fourier basis. This approximation becomes increasingly accurate as the matrix size grows, provided the entries of the Toeplitz matrix decay sufficiently fast away from the main diagonal \citep{Gray_2006_TCM}. In our case, the $\bold{Q}^{i,j}$ matrices are built from cross-correlations of Green functions, whose energy naturally decays with time lag. As a result, the $\bold{Q}^{i,j}$ matrices satisfy the conditions under which the Toeplitz–circulant approximation is valid, and for sufficiently long recording times they can be treated as nearly diagonal in the Fourier domain.
The $2N_t$-length Fourier representation, 
$\bold{\hat{Q}}^{i,j}=\bold{F}\left( \begin{smallmatrix} \bold{{Q}}^{i,j} & \bold{0} \\ \bold{0} & \bold{0} \end{smallmatrix} \right)\bold{F}^T$, is given by
\begin{equation}
\bold{\hat{Q}}^{i,j}_{k,l} = \frac{1}{2N_t}\sum_{m=0}^{N_t-1}\sum_{n=0}^{N_t-1}\bold{Q}^{i,j}_{m,n} e^{\frac{\mathrm{i} 2\pi (km-ln)}{2N_t}}.
\end{equation}
Since the energy of each block $\bold{Q}^{i,j}$ is primarily concentrated near the main diagonal in the Fourier domain, we can threshold the Fourier coefficients based on their magnitude, setting all coefficients below a certain threshold to zero. This yields a sparse approximation of each block and, consequently, the full matrix. The resulting sparse approximation can then be constructed and decomposed using sparse matrix algebra. However, using only the Fourier diagonal elements of each block $\bold{Q}^{i,j}$ provides an accurate approximation, enabling efficient storage of the data-space Hessian matrix by retaining only these components. 

\subsubsection{Computation of the Fourier diagonals}
 The diagonal coefficients of $\bold{\hat{Q}}^{i,j}$ are computed as
\begin{equation}
\bold{\hat{q}}^{i,j}_{k}=
\bold{\hat{Q}}^{i,j}_{k,k} = \frac{1}{2N_t}\sum_{m=0}^{N_t-1}\sum_{n=0}^{N_t-1}\bold{Q}^{i,j}_{m,n} e^{\frac{\mathrm{i}2\pi k(m-n)}{2N_t}}.
\end{equation}
It is not difficult to show that
\begin{equation}
\bold{\hat{q}}^{i,j}_{k}
= \frac{1}{2N_t}\sum_{\xi=-N_t+1}^{N_t-1}\bold{q}^{i,j}_{\xi} e^{\frac{\mathrm{i}2\pi k\xi}{2N_t}},
\end{equation}
where $\bold{q}^{i,j}$ of length $2N_t-1$ is obtained by summing the elements of $\bold{Q}^{i,j}$ along its descending diagonals. Considering the structure of $\bold{Q}^{i,j}$ as defined in \eref{Q}, we get
\begin{equation} 
\bold{q}^{i,j}_{\xi} =\sum_{n-m=\xi} \bold{Q}^{i,j}_{m,n}= \sum_{n-m=\xi}  \bold{W}_{m,n} \bold{G}^{i,j}_{m,n},
\end{equation}
for $\xi=-N_t+1,-N_t+2,...,N_t-1$. 
The weighting function is defined as  
\begin{equation} \label{weightingfunction}
\bold{W}(m,n) = \min(N_t - m,N_t - n), \quad m,n = 0, \dots, N_t - 1.
\end{equation}

\subsubsection{Frequency domain inversion}
The system \eqref{Qsystem}  can be transformed into the frequency domain as:
\begin{equation} \label{Qsystemfreq}
\hat{\bold{Q}} \delta \hat{\bold{d}}^e =  \delta \hat{\bold{d}},
\end{equation}
where $\hat{\bold{Q}}$ is a block matrix with diagonal blocks. The $(i,j)$th block is defined by $\bold{\hat{q}}^{i,j}$ along its main diagonal. The vectors $\delta \hat{\bold{d}}^e$ and $ \delta \hat{\bold{d}}$ are obtained by applying a $2N_t$-point Fourier transform to $\delta {\bold{d}}^e$ and  $\delta {\bold{d}}$ along the time axis.

Using the permutation matrix $\bold{\Pi}$ defined above, we may permute the system in equation \eqref{Qsystemfreq} as
\begin{equation} \label{Qsystemfreq_per}
\bold{\Pi}\hat{\bold{Q}}\bold{\Pi}^T \bold{\Pi} \delta \hat{\bold{d}}^e = \bold{\Pi} \delta \hat{\bold{d}},
\end{equation}
giving us the desired system in the frequency domain
\begin{equation}
\begin{pmatrix}
\hat{\bold{Q}}({\omega_1}) & 0   & \cdots & 0 \\
0 & \hat{\bold{Q}}({\omega_2})   & \cdots & 0 \\
\vdots & \vdots   & \ddots & \vdots \\
0 & 0   & \cdots & \hat{\bold{Q}}({\omega_{N_f}}) \\
\end{pmatrix}
\begin{pmatrix}
\delta \hat{\bold{d}}^e({\omega_1})  \\
\delta \hat{\bold{d}}^e({\omega_2}) \\
\vdots  \\
\delta \hat{\bold{d}}^e({\omega_{N_f}}) \\
\end{pmatrix}
=
\begin{pmatrix}
\delta \hat{\bold{d}}({\omega_1})  \\
\delta \hat{\bold{d}}({\omega_2}) \\
\vdots  \\
\delta \hat{\bold{d}}({\omega_{N_f}}) \\
\end{pmatrix}
\end{equation}
where $\omega$ represents the angular frequency, $N_f$ is the number of frequencies. 
The resulting system is block diagonal and thus can be solved separately for each frequency as
\begin{equation}
\delta \hat{\bold{d}}^e({\omega})=  \hat{\bold{Q}}({\omega})^{-1}  \delta \hat{\bold{d}}({\omega}).
\end{equation}
The coefficient matrices $\hat{\bold{Q}}({\omega})$ of size $N_r\times N_r$ can be efficiently inverted and stored in memory. 
The procedure for computing $\hat{\bold{Q}}(\omega)$ across all frequencies is summarized in Algorithm \ref{alg:Hessian}.

\begin{algorithm}[h!]

\caption{Fourier block-diagonal approximation of the data-space Hessian.}
\label{alg:Hessian}
 Build the weighting function $\bold{W}$ using equation~\eqref{weightingfunction}.\\
 Compute Green functions $\bold{g}^i(t,\bold{x})$ for all active receivers $i = 1, \ldots, N_r$.\\
\For{$i = 1,...,N_r$}{
\For{$j = i,...,N_r$}{
        Form the cross-correlation matrix:
        \[
        \bold{G}^{i,j} = \sum_{\bold{x}} 
        \bold{g}^i(:, \bold{x}) 
        \bold{g}^j(:, \bold{x})^T.
        \]\\
        Compute the weighted correlation vector:
        \[
        \bold{q}^{i,j} = 
        \text{summing elements of~} 
        \big(\bold{W} \circ \bold{G}^{i,j}\big) \text{~along its descending diagonals}.
        \]\\
        Compute the frequency-domain representation:\\
        \[
        \hat{\bold{q}}^{i,j}(\omega) 
        = 2N_t\text{-length FFT of~} \bold{q}^{i,j}.
        \]\\
        For each desired frequency $\omega$, set $\hat{\bold{q}}^{i,j}(\omega)$ as the $(i,j)$ entry of $\hat{\bold{Q}}(\omega)$.
         Set $\hat{\bold{q}}^{j,i}(\omega) = 
    \text{conj}({\hat{\bold{q}}^{i,j}(\omega)})$ as the $(j,i)$ entry of $\hat{\bold{Q}}(\omega)$. 
}}
Invert each frequency-domain matrix $\hat{\bold{Q}}(\omega)$.
\end{algorithm}

\begin{figure}[htb!] 
\begin{center}
\includegraphics[width=0.22\linewidth]{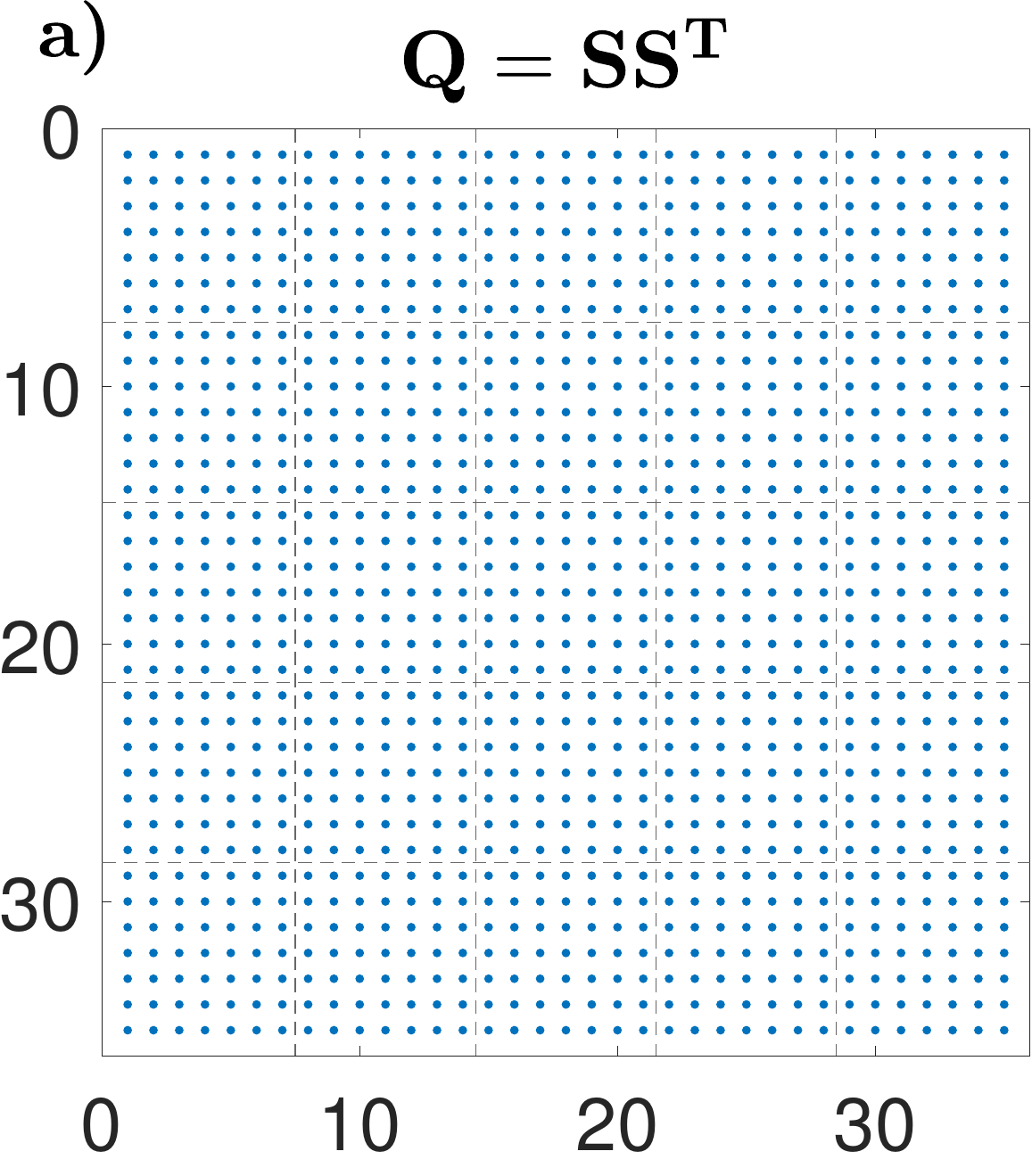}\quad
\includegraphics[width=0.22\linewidth]{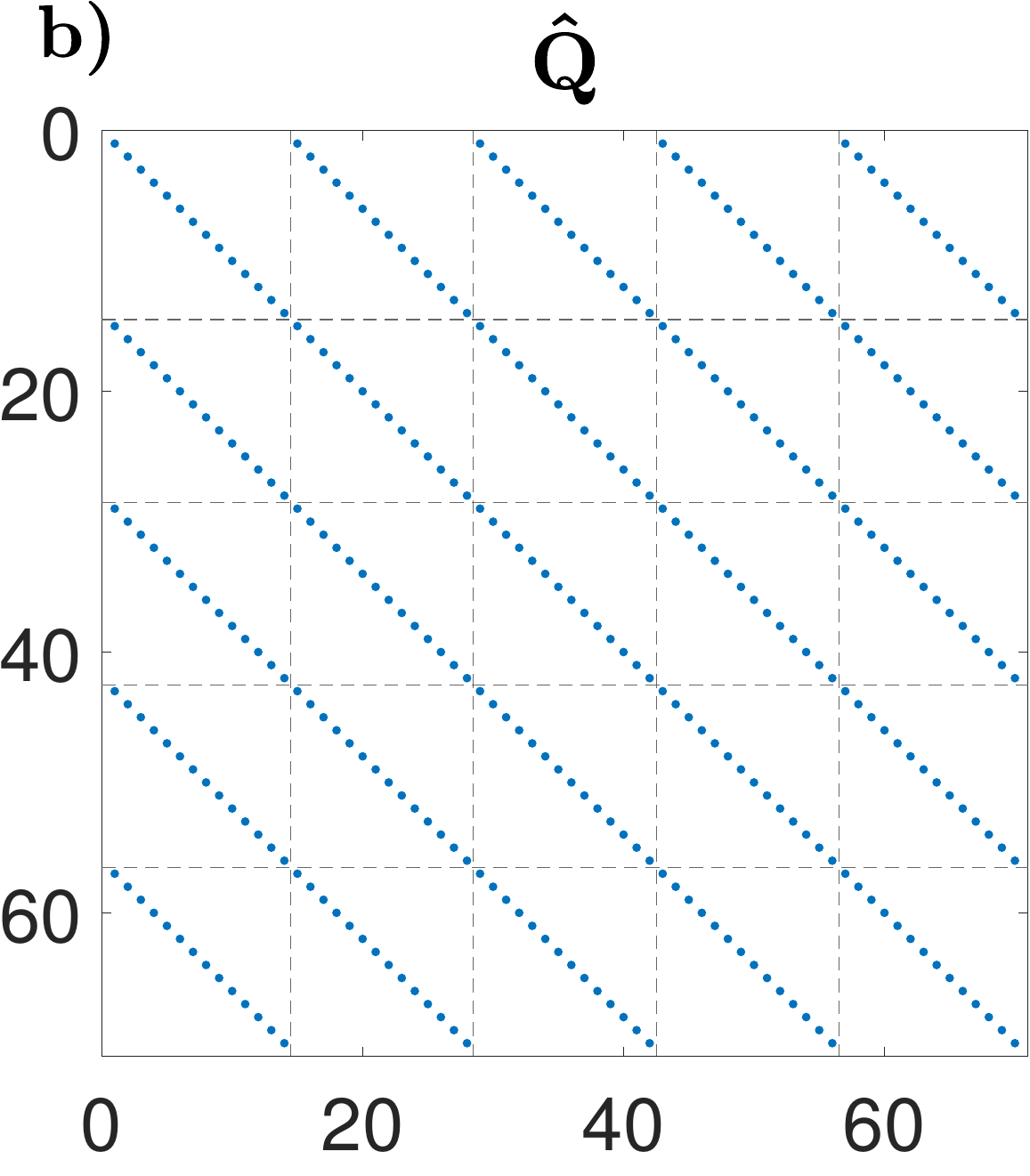}\quad
\includegraphics[width=0.22\linewidth]{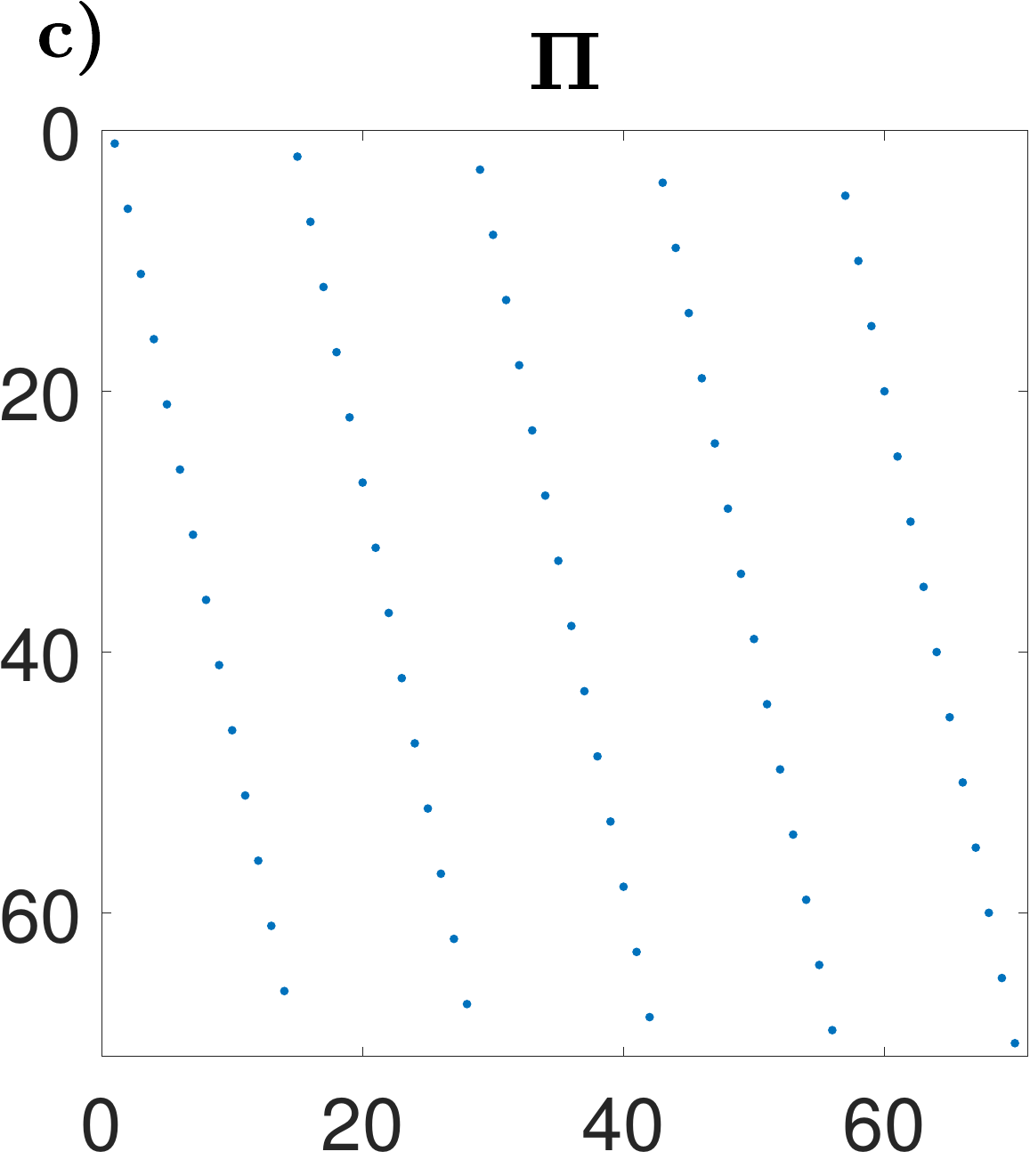}\quad
\includegraphics[width=0.22\linewidth]{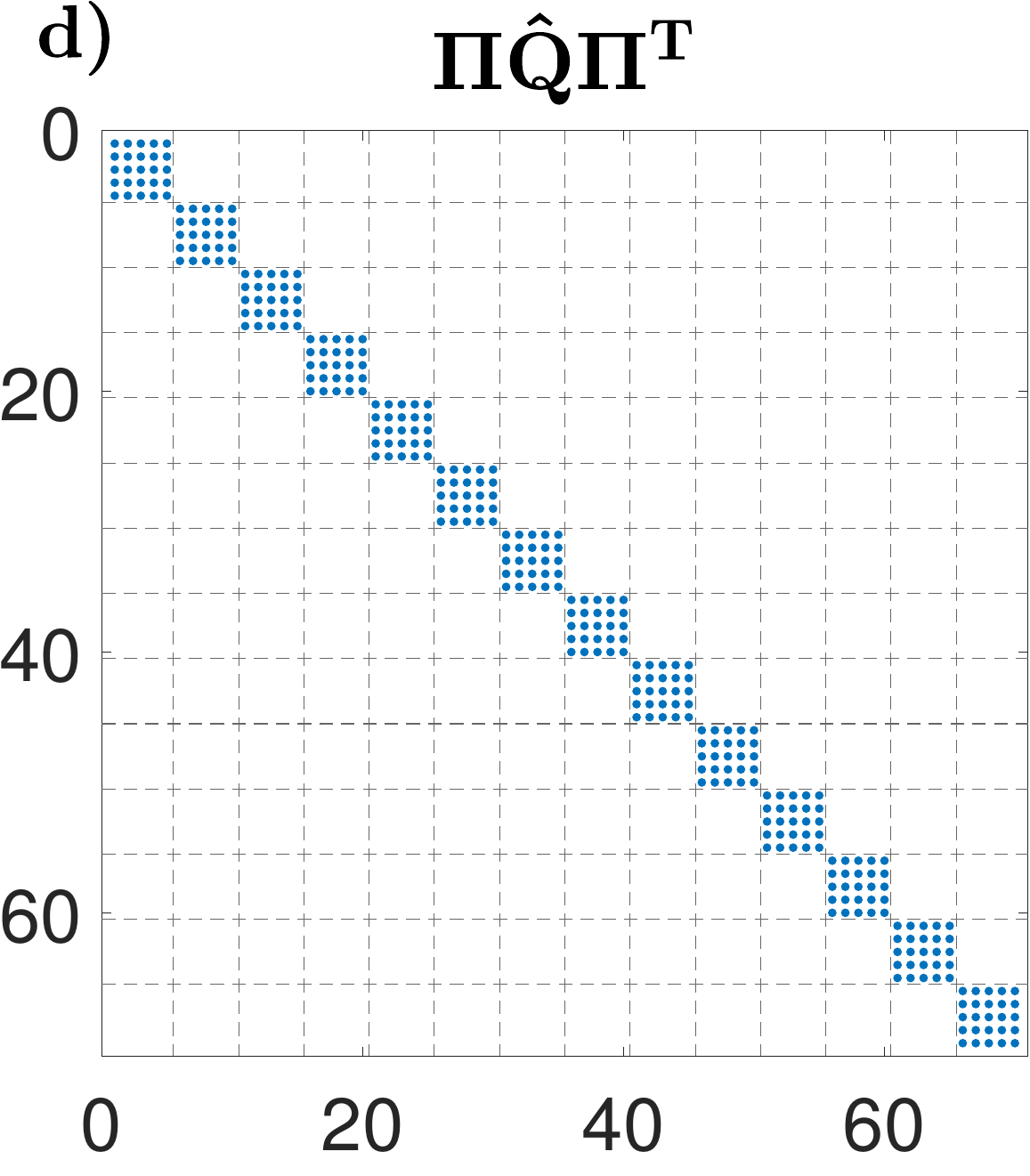}
\caption{Structure of the data-space Hessian matrix for $N_t=7$ and $N_r=5$. (a) Time domain matrix $\bold{Q}$. (b) Frequency domain matrix $\hat{\bold{Q}}$ (for $N_f=2N_t$). (c) Permutation matrix $\bold{\Pi}$. (d) Frequency domain matrix after permutation $\bold{\Pi}\hat{\bold{Q}}\bold{\Pi}^T$. 
}
\label{fig:Qmtx}
\end{center}
\end{figure}

The proposed strategy for direct inversion of the Hessian is illustrated schematically in Figure \ref{fig:Qmtx}. For the case of $N_t = 7$ and $N_r = 5$, the matrix $\bold{Q}$ is a dense $35 \times 35$ matrix composed of $5 \times 5$ blocks, each of size $7 \times 7$ (Figure \ref{fig:Qmtx}a). In the Fourier domain, each of these blocks becomes diagonal (Figure \ref{fig:Qmtx}b).  The permutation matrix $\bold{\Pi}$ (Figure \ref{fig:Qmtx}c) is then used to reorder the rows and columns of $\hat{\bold{Q}}$, transforming it into a block-diagonal matrix (Figure \ref{fig:Qmtx}d), where each block is of size $5 \times 5$.

\subsection{Computational cost}
Direct computation of the Hessian using equation \eqref{Hij} includes multiplying two large Toeplitz matrices of the size $N_t \times N_t$, which costs $\mathcal{O}(N_t^3)$, and summation over $N_x$ samples to make one block of the Hessian. To calculate a complete Hessian, it is necessary to repeat this multiplication for all combinations of $N_r$ receivers, resulting in calculations of the order $N_r^2$. So the total order of calculations will be $\mathcal{O}(N_x N_r^2 N_t^3)$ and the total memory to store the Hessian is $\mathcal{O}(N_r^2 N_t^2)$.
This is while usually $N_x\gg N_t\gg N_r$.
In the proposed {FFT-based method, by using the properties of Toeplitz and Circulant matrices, the order of calculation will be $\mathcal{O}(N_x N_r^2 N_t^2)$, which is $N_t$ times lower than computing using the direct method. 
The required memory to store the Fourier block diagonal approximation of the Hessian is $\mathcal{O}(N_r^2N_{\omega})$.

This cost accounts for building the Hessian using precomputed Green functions, which requires solving one wave equation per receiver. Therefore, constructing the full Hessian with all $N_r$ receivers entails $N_r$ PDE solves for computing the Green functions, followed by Hessian assembly. In the next subsection, we introduce a randomized receiver sampling strategy that reduces the number of active receivers per iteration, thereby lowering the number of required PDE solves and overall computational cost.
\subsubsection{Randomized receiver inversion} \label{RRI}
The random matrix sketching/encoding scheme \citep{Aghazade_2021_RSE} may be employed to significantly decrease the computational and storage costs of building and inverting the Hessian matrix. Basically, the sketching can be applied along the time, space, and receiver axes.  In this study, we resample the time and space axes from the resolution required by finite-difference modeling to the Nyquist rate and reduce the number of receivers using randomized sketching techniques. Specifically, in the numerical examples presented, we use 100 receivers per iteration. To ensure adequate surface coverage, the receiver axis is divided into 100 subsets, and one receiver is randomly selected from each subset at every iteration. Also, in the numerical examples section, we evaluate and compare the impact of varying the number of receivers, from using all receivers to as few as 25.  A comparison of the numerical complexity of building the Hessian matrix in the standard and proposed approaches is presented in Table \ref{tab.num} for different models.

Randomized receiver encoding introduces inherent stochasticity into the inversion process. Different runs, using different random seeds, may produce slightly different recovered models. In this framework, the objective is not to reproduce identical results on a pixel-by-pixel basis, as would be expected in a fully deterministic inversion (i.e., using all receivers simultaneously). Instead, the focus is on ensuring statistical stability and reproducibility of the inversion’s key features and overall quality. The strategy adopted here follows the principles of standard source-encoding approaches \citep{Castellanos_2015_FFW1}. At each iteration, a new set of random encoding weights is generated for the receivers. This frequent regeneration of encoding patterns effectively averages out the crosstalk noise, driving its expectation toward zero and improving the robustness of the inversion.

\begin{table}[!h]
\centering
\caption{Computational complexity and speedup of the Hessian implementation for the Marmousi II and 2004 BP salt model.}
\label{tab.num}
\footnotesize
\begin{tabular}{|c|c|c|c|c|}
\hline
Model       & \begin{tabular}[c]{@{}c@{}}Direct computation\\  $\mathcal{O}(N_x N_r^2 N_t^3)$\end{tabular} & \begin{tabular}[c]{@{}c@{}}Proposed approach\\  $\mathcal{O}(N_x N_r^2  N_t^2)$\end{tabular} & \begin{tabular}[c]{@{}c@{}c@{}}Proposed approach \\ with randomized \\receivers 
 \end{tabular} & Speedup \\ 
\hline
Marmousi II & \begin{tabular}[c]{@{}c@{}} $(172 \times 605) \times 273^2 \times 4334^3$\\  $\simeq 6.31\times 10^{20}$\end{tabular} & \begin{tabular}[c]{@{}c@{}} $(172 \times 605) \times 273^2 \times 4334^2$\\  $ \simeq 1.45\times 10^{17}$\end{tabular}& \begin{tabular}[c]{@{}c@{}} $(57 \times 201) \times 100^2 \times 434^2$\\  $ \simeq 2.16\times 10^{13}$\end{tabular} & $2.9\times10^{7}$\\ \hline
2004 BP&  \begin{tabular}[c]{@{}c@{}} $(220 \times 960) \times 450^2 \times 2813^3$\\  $\simeq 9.52\times 10^{20}$\end{tabular} & \begin{tabular}[c]{@{}c@{}} $(220 \times 960) \times 450^2 \times 2813^2$\\  $\simeq 3.38\times 10^{17}$\end{tabular} & \begin{tabular}[c]{@{}c@{}} $(110 \times 480) \times 100^2 \times 284^3$\\  $\simeq 4.25\times 10^{13}$\end{tabular} & $2.2\times10^{7}$\\ \hline
\end{tabular}
\end{table}

\begin{table}[]
\centering
\caption{Number of PDE solves and average CPU time required for Hessian inversion per iteration using CG, GMRES, and the proposed direct Hessian inversion across three benchmark models. Dashes indicate cases where CG/GMRES were not implemented.}
\label{Table_2}
\begin{tabular}{|c|c|c|ccc|l}
\cline{1-6}
\multirow{2}{*}{Model}                                                     & \multirow{2}{*}{\begin{tabular}[c]{@{}c@{}}Number \\ of sources\end{tabular}} & \multirow{2}{*}{\begin{tabular}[c]{@{}c@{}}Performance\\ metric\end{tabular}} & \multicolumn{3}{c|}{Method}                                                                                         &                      \\ \cline{4-6}
                                                                           &                                                                               &                                                                               & \multicolumn{1}{c|}{CG}   & \multicolumn{1}{c|}{GMRES} & \begin{tabular}[c]{@{}c@{}}Direct\\ inversion\end{tabular} &                      \\ \cline{1-6}
\multirow{2}{*}{Marmousi II}                                               & \multirow{2}{*}{68}                                                           & PDE solutions                                                                 & \multicolumn{1}{c|}{9656} & \multicolumn{1}{c|}{4760}  & 100                                                        &                      \\ \cline{3-6}
                                                                           &                                                                               & CPU time (s)                                                                  & \multicolumn{1}{c|}{---}  & \multicolumn{1}{c|}{---}   & 367                                                        &                      \\ \cline{1-6}
\multirow{2}{*}{2004 BP}                                                   & \multirow{2}{*}{67}                                                           & PDE solutions                                                                 & \multicolumn{1}{c|}{9916} & \multicolumn{1}{c|}{4824}  & 100                                                        &                      \\ \cline{3-6}
                                                                           &                                                                               & CPU time (s)                                                                  & \multicolumn{1}{c|}{---}  & \multicolumn{1}{c|}{---}   & 607                                                        &                      \\ \cline{1-6}
\multirow{2}{*}{\begin{tabular}[c]{@{}c@{}}Cropped\\ 2004 BP\end{tabular}} & \multirow{2}{*}{30}                                                           & PDE solutions                                                                 & \multicolumn{1}{c|}{6000} & \multicolumn{1}{c|}{3000}  & 100                                                        &                      \\ \cline{3-6}
                                                                           &                                                                               & CPU time (s)                                                                  & \multicolumn{1}{c|}{2205} & \multicolumn{1}{c|}{1092}  & 51                                                         & \multicolumn{1}{c}{} \\ \cline{1-6}
\end{tabular}
\end{table}

\section{Numerical examples} \label{NumEx}
In this section, we demonstrate the effectiveness of the proposed direct Hessian inversion method using two synthetic examples from the Marmousi II and 2004 BP salt benchmark models. For each example, the parameter $\mu$ in equation \ref{Qmatrix} was chosen to be a fraction ($10^{-2}-10^{-4}$) of the maximum eigenvalue of $\bold{SS}^T$. Additionally, at each iteration, the projection operator $\text{proj}_\mathcal{C}$ is defined by the true minimum and maximum velocity values, and the model parameters corresponding to the water layer are assumed to be known. Also, to make the algorithms more stable, specifically at early iterations, we apply a simple damping to the multiplier updates as proposed by \citep[][ their Equation 27 for $k_0=5$]{Gholami_2023_MWI}.

We compare three approaches: (i)  reduced FWI (or simply FWI), (ii) multiplier-based FWI using the diagonal Hessian approximation of \citep{Gholami_2022_EFW}, and (iii) multiplier-based FWI using the proposed direct Hessian inversion method. The FWI is implemented using Algorithm 1 with the multiplier $\bold{\Delta d}_s$ set to zero and with no source extension on the right-hand side of the wave equation, i.e., using the standard forward wavefield $\bu_s = \bA(\bm_k)^{-1}\bb_s$ in place of the extended wavefield $\bu^e_s$.


\subsection{Marmousi II}
The model,  spanning 17 km in offset and 3.5 km in depth (Figure \ref{MARM_results}a), is discretized with a grid spacing of 31.25 m. The acquisition setup consists of 273 surface sources (spaced every 62.5 m) and 68 seabed receivers (spaced every 250 m), with a recording time window of 13 s and a sampling interval of 3 ms. Using reciprocity, sources were treated as receivers and vice versa. A 6 Hz Ricker wavelet, bandpass filtered between 2 Hz and 12 Hz, was used as the source.  
%
To improve computational efficiency, data and Green functions were resampled along both the time and space axes to the source Nyquist rate (leading to 10 times undersampling in time and $9=3\times 3$ in space). Additionally, at each iteration, only 100 out of the 273 receivers were randomly selected for processing. Since the computational complexity of Hessian construction scales with the square of the number of receivers, this random selection led to a computational gain of approximately $(\frac{273}{100})^2 = 7.45$. 
The computational cost of Hessian construction per iteration for the Marmousi II model is summarized in Table \ref{tab.num}.

\subsubsection{Hessian accuracy test}
To assess the accuracy of Hessian inversion using the direct frequency-domain approach, we compare the data predicted by the inverted extended source with the observed data for a source located at an 8.5 km offset. Specifically, deblurring the data residuals using the data-space Hessian, $\bold{Q}^{-1}\delta\bold{d}$ (equation \ref{Qsystem}), yields the receiver-side Lagrange multipliers, $\delta\bold{d}^e$. Back-propagating these multipliers as adjoint source provides the back-propagating wavefield, $\delta\bold{b} =  \bold{S}^T\delta\bold{d}^e$. These wavefields act as source-side multipliers and serve as the source extension, forming the extended source $\bold{b}+\delta\bold{b}$.

From this extended source, the extended wavefield is computed as $\bu^e= \bA(\bm_k)^{-1}(\bb + \delta \bb)$ where $\bm_k$ is the current model.
Sampling this wavefield at the receiver locations gives the predicted data: $\bold{d}^{\text{pre}} =\bP\bu^e(\delta\bold{d}^e)= \bold{S}[\bold{b} + \bold{S}^T\delta\bold{d}^e]$. The relative error between the observed and predicted data is then defined as
\begin{equation} 
   \text{Relative Error}= \frac{\|\bd-\bP \bu^e(\delta\bold{d}^e)]\|_2^2}{\|\bd\|_2^2}.
\end{equation}
For a small damping parameter $\mu$ regularizing the Hessian, these predicted data should match the observed data regardless of the accuracy of $\bm_k$. 
Importantly, when the linear system $\bold{Q}\delta\bold{d}^e=\delta\bold{d}$ is solved exactly, the relative error becomes fixed and independent of the inversion method. Conversely, any deviation from this solution reflects the accuracy of the Hessian inversion.
For instance, if the system is solved exactly so that $\delta\bold{d}^e=\bold{Q}^{-1}\delta\bold{d}$, then substituting this into $\bu^e(\delta\bold{d}^e)$ yields, after simplification, $\bd-\bP\bu^e(\delta\bold{d}^e) = \mu\delta\bold{d}^e$ \citep{Gholami_2022_EFW}.
However, when iterative solvers such as CG or GMRES \citep{Nocedal_2006_NO} are used to approximate the solution to $\bold{Q}\delta\bold{d}^e=\delta\bold{d}$, it is still possible to achieve the same Relative Error, but the quality of the residual $\bd-\bP\bu^e(\delta\bold{d}^e)$ can vary. This occurs because any approximation in the Hessian inversion propagates back through the medium as part of the estimated adjoint source. These inaccuracies act as spurious sources, introducing unwanted energy into the modeled wavefield that is recorded at the receivers, thereby degrading the data fit. 


We compare the performance of the direct inversion method with CG and GMRES, requiring two PDE solves per iteration. The MATLAB built-in functions were used to implement these iterative methods. In each case, iterations were stopped once the difference between the predicted and observed data reached the fit level achieved by the direct method.

Figures \ref{figmarmdata}a and b show the recorded data from the true model (Figure \ref{MARM_results}a) and the initial model (Figure \ref{MARM_results}b), respectively, for a 4 Hz Ricker wavelet. Figures \ref{figmarmdata}d-f show the predicted data using the proposed direct inversion, CG, and GMRES methods, respectively, and figures \ref{figmarmdata}g-i show the difference between each of them with the data in the true model (\fref{figmarmdata}a). Figure \ref{figmarmdata}c shows the Relative Error between the observed and predicted data for these methods. To match the accuracy of the direct inversion, the CG and GMRES methods required 71 and 35 iterations, which means 142 and 70 PDE solves, respectively. Also, for different sources, almost the same number of iterations has to be repeated. Meanwhile, direct inversion requires only 100 PDE solves and can be used for all sources.

From Figures \ref{figmarmdata}g–i we can see that the proposed direct Hessian inversion reproduces the weak, late-arriving events particularly well, though some residual energy remains for the strong direct and diving waves. The CG method, owing to its greedy residual-minimization nature, fits the strong early-arrival events more accurately but struggles to match the weaker late arrivals. GMRES provides an intermediate behavior: it yields a better fit for strong arrivals than the direct inversion but still fails to capture the late arrivals as effectively.

\subsubsection{Inversion}
The inversion was initialized from the 1D starting model (Figure \ref{MARM_results}b) and performed in a single inversion cycle using a multiscale strategy \citep{Boonyasiriwat_2009_EMM}. Data were first inverted in the 2–6 Hz band (peak frequency 3 Hz) over 280 iterations. The resulting model was then used as the starting point for an additional 100 iterations in the full 2–12 Hz band (peak frequency 6 Hz). Figures \ref{MARM_results}c–e show the models recovered using reduced FWI, multiplier-based FWI with the diagonal Hessian approximation, and the proposed direct Hessian inversion method, respectively. Reduced FWI failed to converge and became trapped in a local minimum, as evidenced by its nearly flat convergence curve during the second frequency band (green curve in Figure \ref{MARM_curves}). In contrast, both multiplier-based approaches successfully mitigated cycle-skipping and produced geologically consistent results. The model obtained with the direct Hessian inversion shows noticeably improved accuracy, particularly in the deeper sections, as confirmed by the vertical velocity profiles in Figures \ref{MARM_log}a and \ref{MARM_log}b.
The presence of a low-velocity anomaly at a depth of 1 km and a distance of 3 km posed a challenge for reconstructing the velocity model beneath it. This limitation can be alleviated by performing a second cycle of inversion.
Figures \ref{MARM_data_grad}a–d display seismograms computed using the initial model and the models obtained by the three inversion methods. Their differences with respect to the true seismograms are shown in Figures \ref{MARM_data_grad}e–h. These comparisons clearly demonstrate that, under identical inversion settings, implementing the Hessian produces velocity models that fit the data more accurately across all times and offsets.

To evaluate the impact of randomized receiver selection on inversion quality, we conducted a series of experiments with different numbers of receivers. Inversions were performed with all 273 receivers, and 200, 100, 50, and 25 randomly selected receivers per iteration. Figure \ref{MARM_data_spec}a–e displays a shot gather in the time domain for each case: (a) all 273 receivers, (b) 200 receivers, (c) 100 receivers, (d) 50 receivers, and (e) 25 receivers. The corresponding frequency–wavenumber spectra are shown in Figure \ref{MARM_data_spec}f–j. As expected, aliasing becomes increasingly evident as the number of receivers decreases. However, due to the randomized nature of the downsampling, the aliasing remains incoherent, satisfying the conditions required for randomized inversion approaches \citep{Ben_2011_EFF}.  The resulting inverted velocity models for 273, 200, 50, and 25 receivers are shown in Figure \ref{MARM_rec_test}a–d, respectively. The result for 100 receivers is presented separately in Figure \ref{MARM_results}e. The convergence of different cases is compared in \fref{MARM_curve_receiver}. Remarkably, even with as few as 25 receivers (Figure \ref{MARM_rec_test}d), a reasonable velocity model is recovered, while achieving a significant reduction in computational cost, by a factor of $\left(\frac{273}{25}\right)^2 = 119.25$. This clearly demonstrates the efficiency and effectiveness of the randomized receiver inversion strategy. 

Finally, the result shown in Figure \ref{MARM_results}e was obtained by updating the Hessian matrix at every iteration. However, such frequent updates may not always be necessary. To investigate the effect of the Hessian update frequency, we repeated the same experiment but updated the Hessian only every 5 and 10 iterations. The resulting inverted velocity models are shown in Figure \ref{MARM_recycle}a and b, with their corresponding convergence curve plotted in Figure \ref{MARM_recycle_curve}. The result remains highly accurate and very similar to the case where the Hessian is updated at every iteration. This experiment demonstrates that the proposed method is robust to moderate inaccuracies in the Hessian, thanks to the error-correcting effect of the Lagrange multipliers. 

\begin{figure}[!h]
    \centering
    \includegraphics[width=1\linewidth]{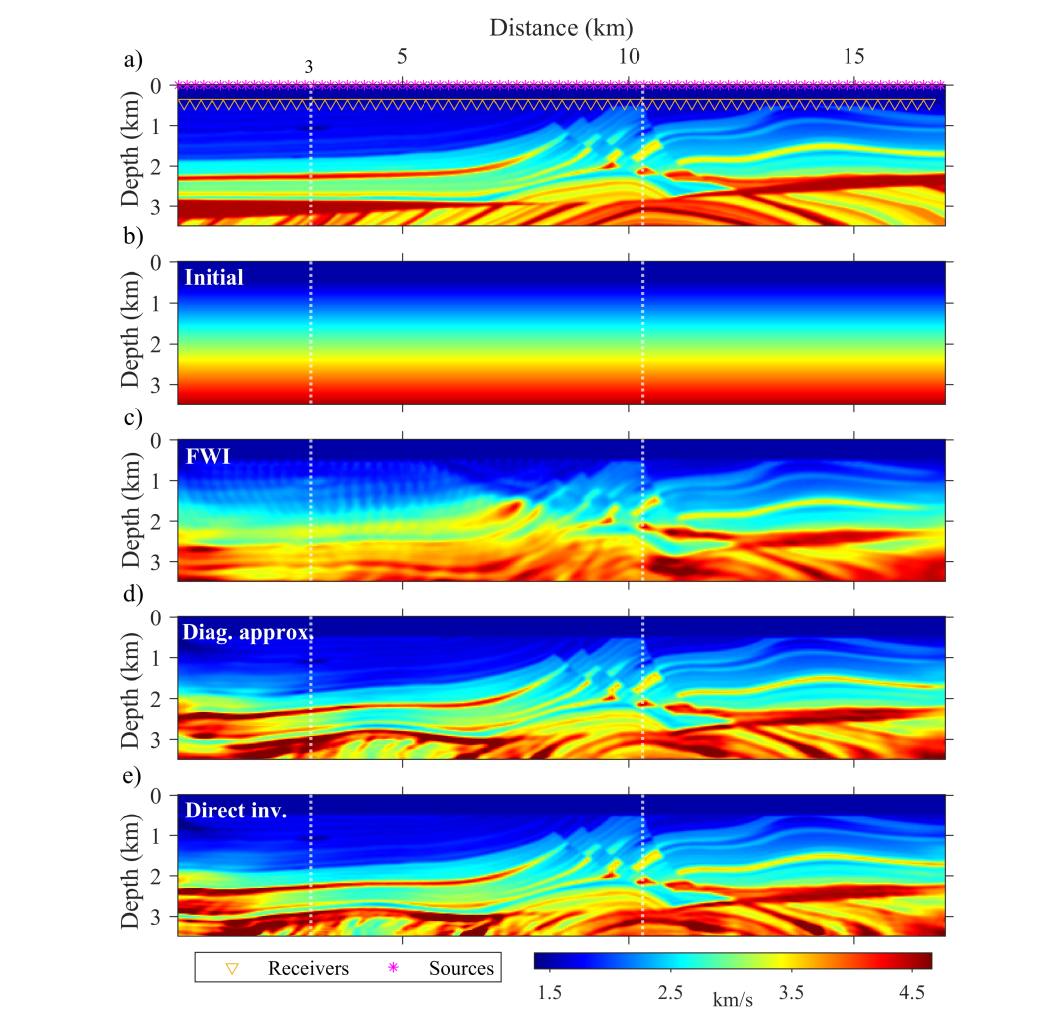}
    \caption{(a) Marmousi II model. (b) Initial model. (c–e) Inversion results using (c) reduced FWI, (d) multiplier-based FWI with diagonal Hessian approximation, and (e) multiplier-based FWI with direct Hessian inversion.}
    \label{MARM_results}
\end{figure}

\begin{figure}[!h]
    \centering
    \includegraphics[width=1\linewidth]{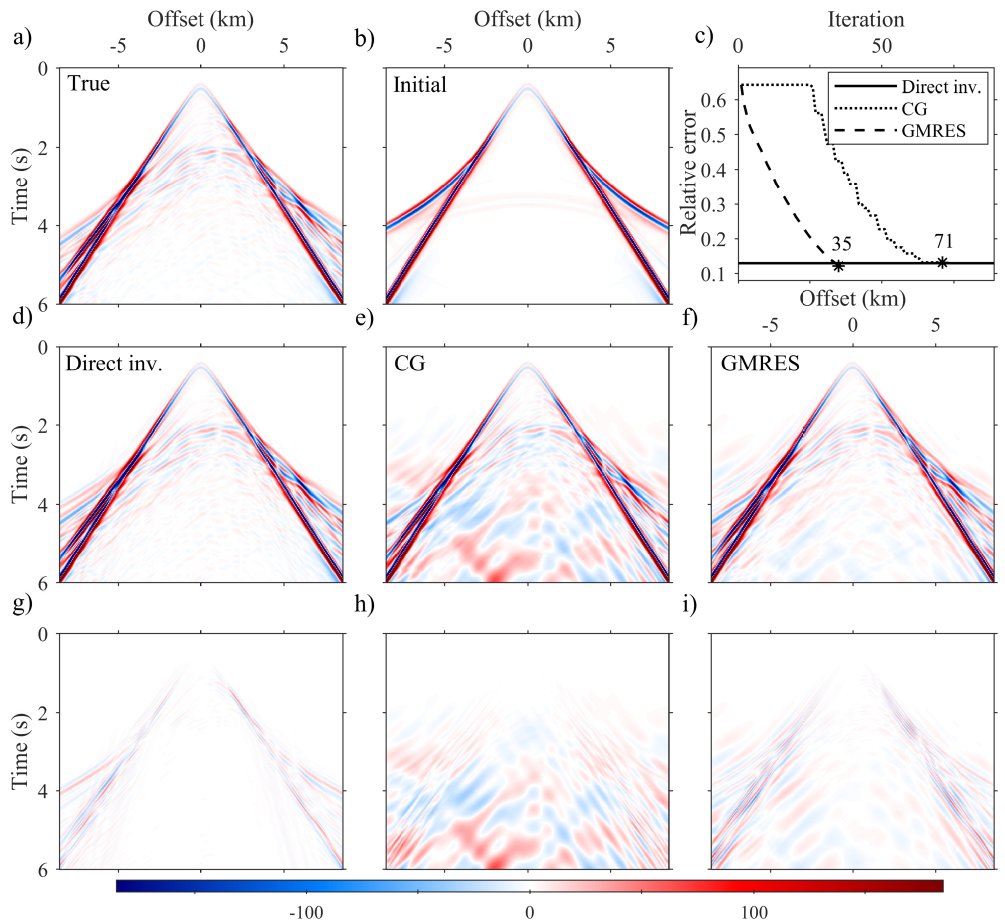}
    \caption{Shot gather comparisons for the Marmousi II model for a source at 8.5 km. (a) True data. (b) Data from initial model. (c) Relative error vs. iteration for direct inversion, CG, and GMRES. (d–f) Predicted data using direct inversion, CG, and GMRES, respectively. (g–i) Residuals relative to true data. 
    }
    \label{figmarmdata}
\end{figure}

\begin{figure}[!htb]
    \centering
    \includegraphics[width=0.5\linewidth]{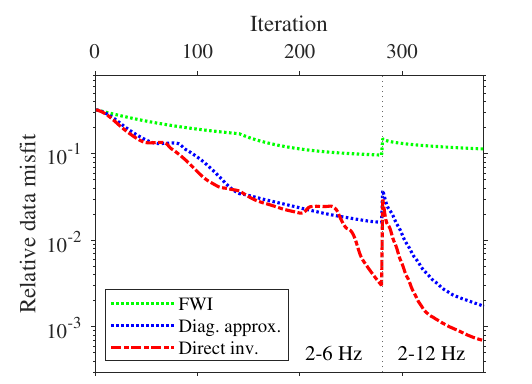}
    \caption{Relative data misfit trajectories for the Marmousi II inversion. Green: reduced FWI, blue: diagonal Hessian approximation, red: proposed direct Hessian inversion.}
    \label{MARM_curves}
\end{figure}

\begin{figure}[!h]
    \centering
    \includegraphics[width=0.4\linewidth]{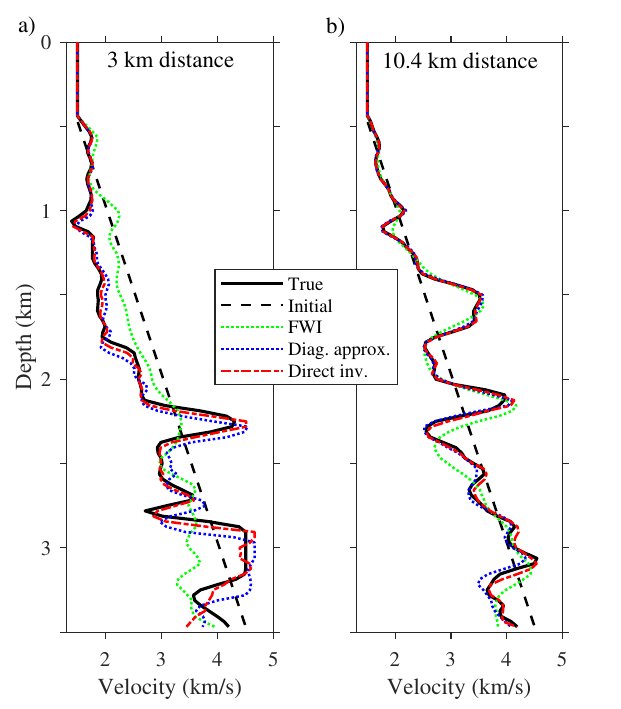}
    \caption{(a–b) Vertical velocity profiles extracted from the true model (black), the initial model (dashed black), and the inverted results obtained using reduced FWI (green) and multiplier-based FWI with a diagonal Hessian approximation (blue), and the proposed direct Hessian inversion (red) at (a) 3 km and (b) 10.4 km horizontal distance.}
    \label{MARM_log}
\end{figure}

\begin{figure}[!htb]
    \centering
    \includegraphics[width=1\linewidth]{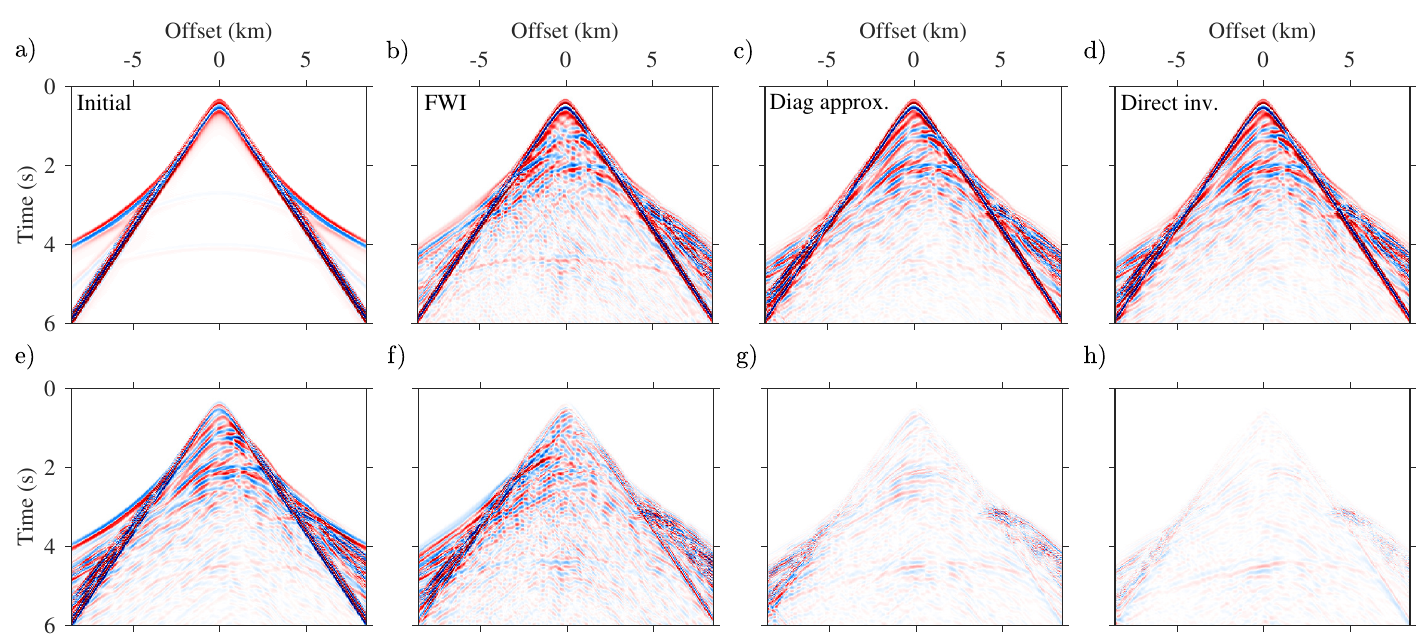}
    \caption{Seismograms computed for the Marmousi II example using (a) the 1D initial model and the inverted models obtained by (b) reduced FWI and multiplier-based FWI with (c) a diagonal Hessian approximation and (d) the proposed direct Hessian inversion. (e–h) Residuals relative to true data. }
    \label{MARM_data_grad}
\end{figure}

\begin{figure}[!htb]
    \centering
    \includegraphics[width=1\linewidth]{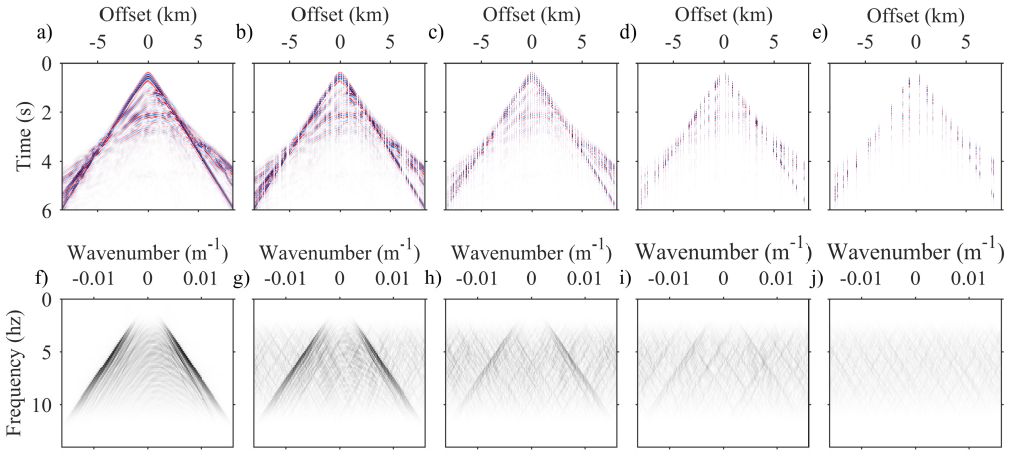}
    \caption{Shot gather of the source at 8.5 km distance computed in the Marmousi II model in the time domain for different numbers of randomly selected receivers: (a) 273 (all), (b) 200, (c) 100, (d) 50, and (e) 25. Panels (f-j) show the corresponding shot gathers after applying 2D Fourier transform for (f) all 273 receivers, (g) 200 receivers, (h) 100 receivers, (i) 50 receivers, and (j) 25 receivers.}
    \label{MARM_data_spec}
\end{figure}

\begin{figure}[!h]
    \centering
    \includegraphics[width=1\linewidth]{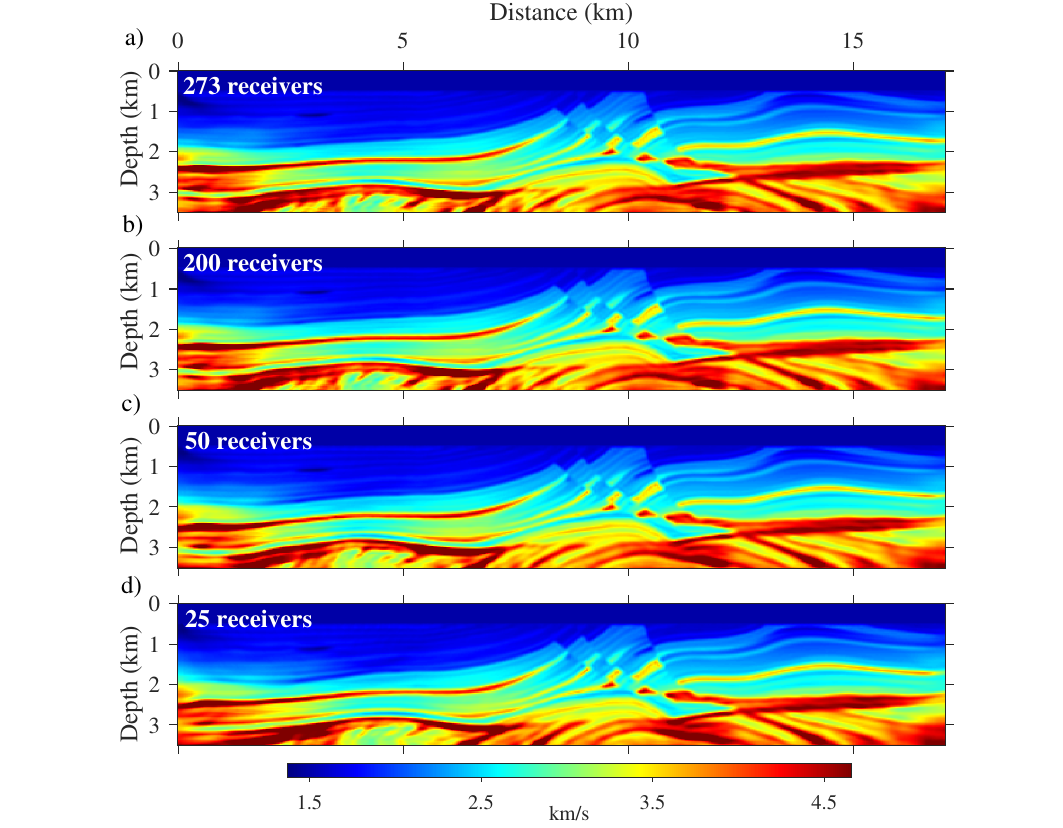}
    \caption{Inverted Marmousi II models with different numbers of randomly selected receivers per iteration: (a) 273 (all), (b) 200, (c) 50, and (d) 25.}
    \label{MARM_rec_test}
\end{figure}

\begin{figure}[!htb]
    \centering
    \includegraphics[width=0.5\linewidth]{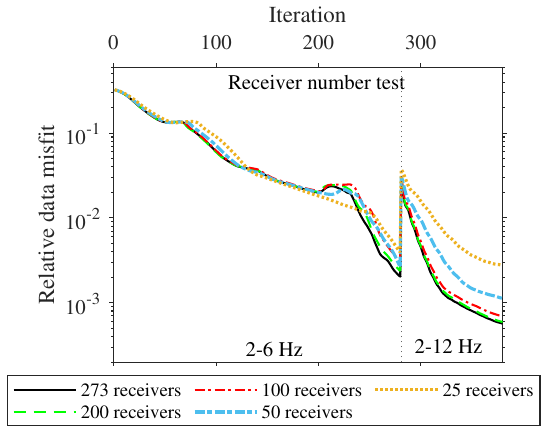}
    \caption{Trajectory of the relative data misfit for the Marmousi II inversion with different numbers of randomly selected receivers per iteration: 273 (all), 200, 100, 50, and 25.}
    \label{MARM_curve_receiver}
\end{figure}

\begin{figure}[!h]
    \centering
    \includegraphics[width=1\linewidth]{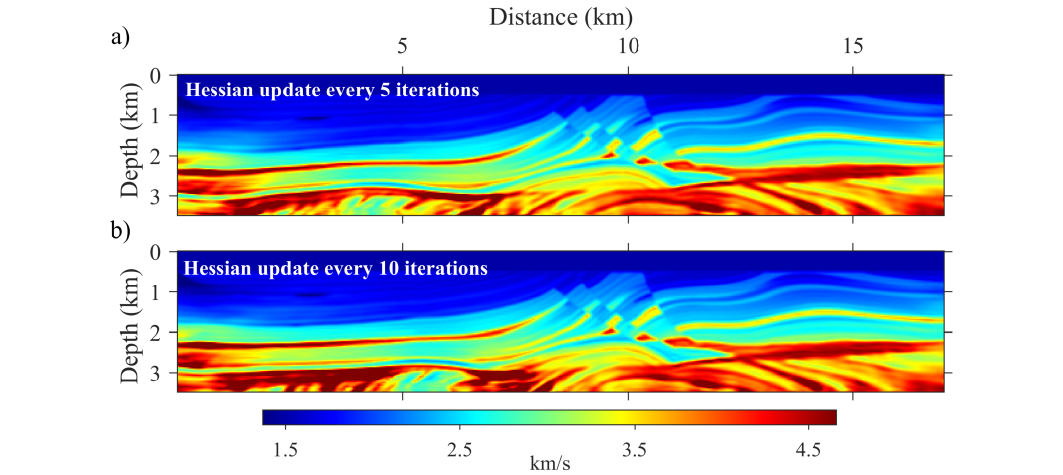}
    \caption{Inverted Marmousi II models obtained with multiplier-based FWI using the proposed direct Hessian inversion, with  updating Hessian every (a) 5 iterations and (b) 10 iterations.}
    \label{MARM_recycle}
\end{figure}

\begin{figure}[!h]
    \centering
    \includegraphics[width=0.5\linewidth]{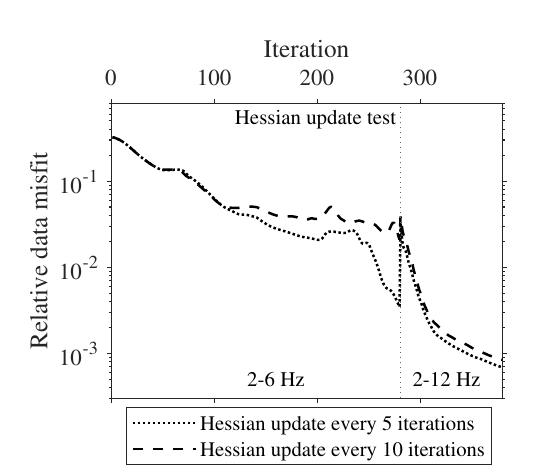}
    \caption{Trajectory of the relative data misfit for the Marmousi II inversion using direct Hessian inversion with updating the Hessian every 5 iterations (dotted line), and every 10 iterations (dashed line).}
    \label{MARM_recycle_curve}
\end{figure}



\subsection{2004 BP salt model}
The model,  spanning 67.5 km in offset and 12 km in depth (Figure \ref{BP_results}a), is discretized with a grid spacing of 75 m. The acquisition setup consists of 450 surface sources (spaced every 150 m) and 67 seabed receivers (spaced every 1000 m), with a recording time window of 22.5 s and a sampling interval of 8 ms. Using reciprocity, sources were treated as receivers and vice versa. A 5 Hz Ricker wavelet, bandpass filtered between 1.5 Hz and 4.5 Hz, was used as the source. 

To improve computational efficiency, data and Green functions were resampled along both the time and space axes to the Nyquist rate (leading to 10 times undersampling in time and $4=2\times 2$ times in space). Additionally, at each iteration, only 100 out of the 450 receivers were randomly selected for processing. Since the computational complexity of Hessian construction scales with the square of the number of receivers, this random selection led to a computational gain of approximately $(\frac{450}{100})^2 = 20.25$. 
The computational cost of Hessian construction per iteration for the 2004 BP salt model is summarized in Table \ref{tab.num}.

\begin{figure}[!h]
    \centering
    \includegraphics[width=1\linewidth]{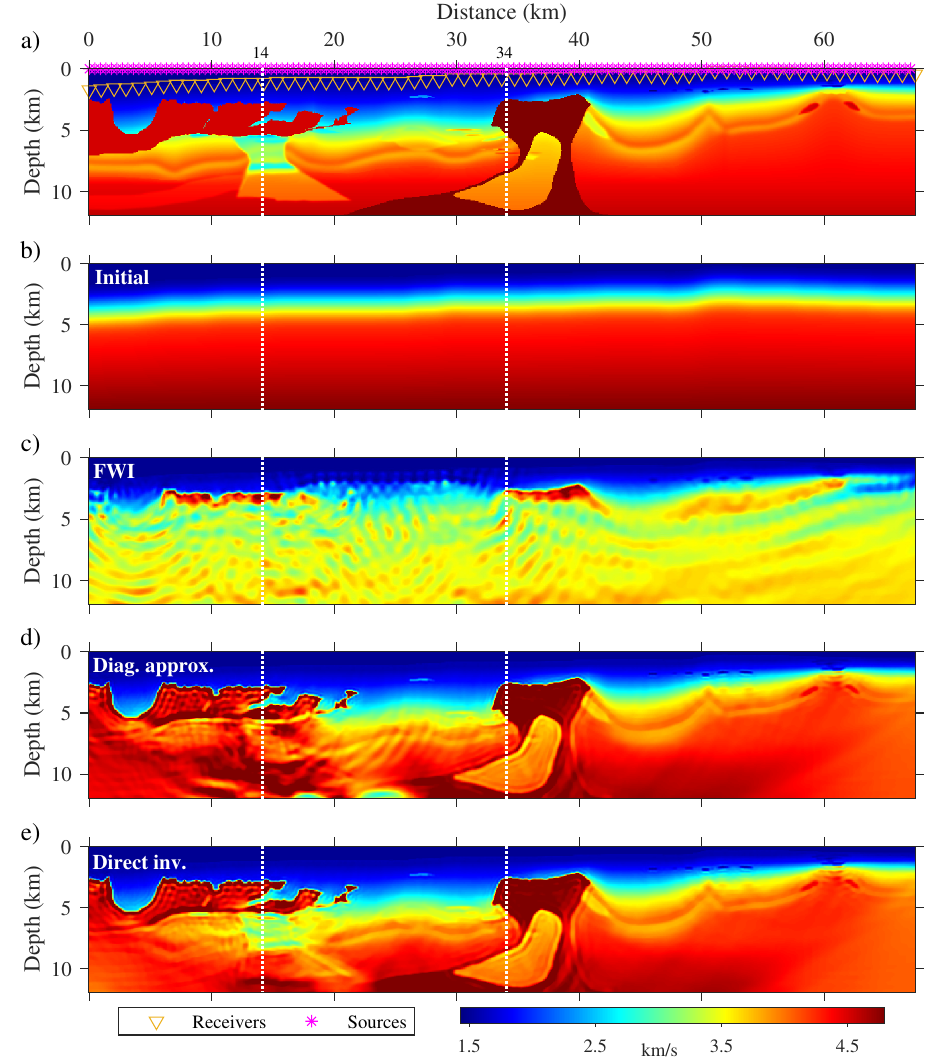}
    \caption{(a) 2004 BP salt model. (b) Initial model. Reconstructed model using (c) reduced FWI, and using multiplier-based FWI with (d) diagonal Hessian approximation and (e) direct Hessian inversion.}
    \label{BP_results}
\end{figure}

\subsubsection{Hessian accuracy test}
As with the Marmousi II model, we first test the accuracy of the proposed approach by calculating predicted data using the initial model and comparing it with the data acquired from the true model. Figures \ref{figbpdata}a and \ref{figbpdata}b show the data corresponding to the source at a 34 km for a 2 Hz Ricker wavelet, computed using the true model (Figure \ref{BP_results}a) and the initial model (Figure \ref{BP_results}b), respectively. Figures \ref{figbpdata}d-f show the predicted data using the proposed direct inversion, CG, and GMRES methods, respectively, and \fref{figbpdata}g-i show the difference between each of them with the data in the true model (\fref{figbpdata}a). The Relative Error curves versus iteration alongside the error of the direct method are shown in Figure \ref{figbpdata}c. To match the accuracy of the direct inversion, the CG and GMRES methods required 74 and 36 iterations, which means 148 and 72 PDE solves, respectively. Also, for different sources, almost the same number of iterations has to be repeated. Meanwhile, direct inversion requires only 100 PDE solves and can be used for all sources.

Inspection of the residual gathers in Figures \ref{figbpdata}g–i reveals that the proposed direct Hessian inversion reproduces the weak, late-arriving events particularly well, though some residual energy remains for the strong direct and diving waves. The CG method, owing to its greedy residual-minimization nature, fits the strong early-arrival events more accurately but struggles to match the weaker late arrivals. GMRES provides an intermediate behavior: it yields a better fit for strong arrivals than the direct inversion but still fails to capture the late arrivals as effectively.

\begin{figure}[!h]
    \centering
    \includegraphics[width=1\linewidth]{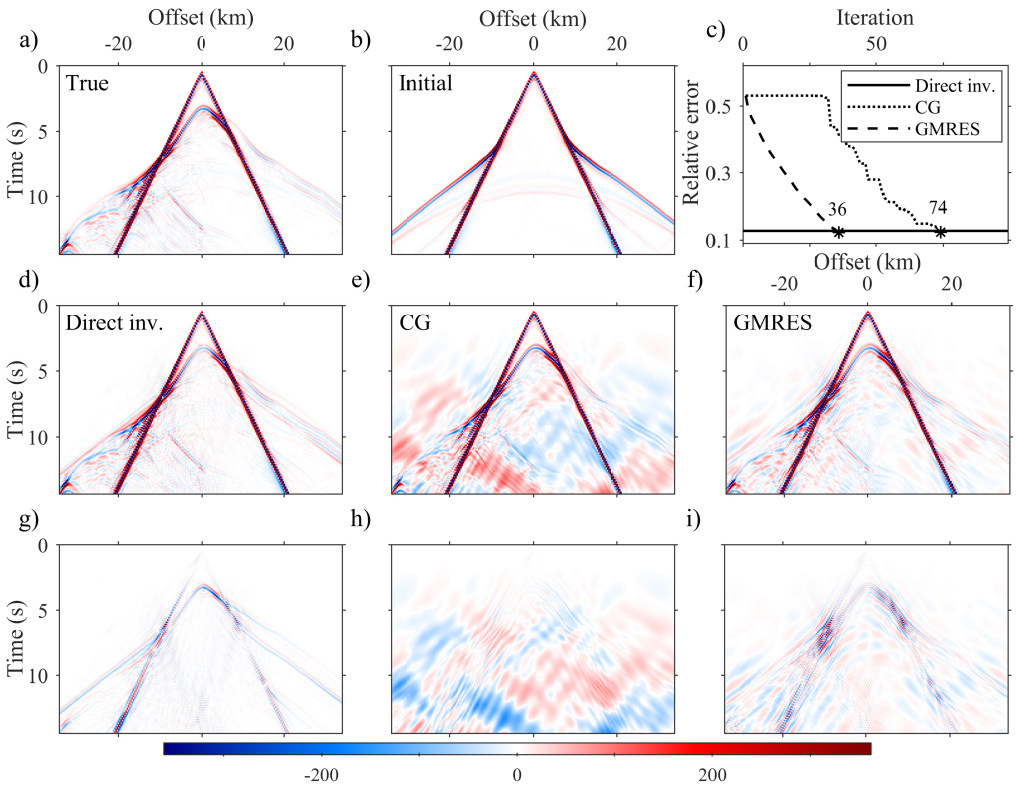}
    \caption{Shot gather for the source at 34 km distance computed using (a) the true 2004 BP salt model (Figure \ref{BP_results}a) and (b) the initial model (Figure \ref{BP_results}b). (c) The Relative Error between observed and predicted data for CG and GMRES methods compared with the direct inversion (solid horizontal black line). (d-f) Predicted shot gather for the same source computed in the initial model using the (d) direct Hessian inversion, (e) CG iteration, and  (f) GMRES iteration. (g–i) Residuals relative to true data.}
    \label{figbpdata}
\end{figure}

\subsubsection{Inversion}
We conducted one inversion cycle, involving a multiscale approach \citep{Boonyasiriwat_2009_EMM}. Also, we use two different grid sizes, 150 m and 75 m grid spacing on each step of the multiscale approach, satisfying 5 grid points per minimum wavelength \citep{Gholami_2023_MWI}. In the first stage, data in the 1.5–1.7 Hz band were inverted over 250 iterations using the 150 m grid. This result served as the starting model for the second stage, where 100 additional iterations were performed in the broader 1.5–4.5 Hz band with a peak frequency of 3.5 Hz on the finer 75 m grid.

Figures \ref{BP_results}c–e show the inverted models obtained using reduced FWI, and the multiplier-based FWI with the diagonal Hessian approximation, and the proposed direct Hessian inversion method, respectively. Reduced FWI failed to converge and became trapped in a local minimum, as evidenced by its nearly flat convergence curve (green curve in Figure \ref{BP_curve}). In contrast, both multiplier-based approaches successfully mitigated the cycle-skipping problem and converged to geologically plausible solutions. The model obtained with the proposed direct Hessian inversion exhibits noticeably improved accuracy, particularly in the left portion and in the subsalt region. Similar to the Marmousi model, in this model also the presence of a low-velocity anomaly at a depth of 1.8 km and a distance of 48-54 km posed a challenge for reconstructing the velocity model beneath it; this can be improved by performing a second cycle of inversion.
Figures \ref{BP_log}a and \ref{BP_log}b compare the true, initial, and inverted velocity profiles at two spatial locations (14 km and 34 km). These comparisons clearly demonstrate that incorporating the Hessian improves the reconstruction of deep velocity structures, albeit at the expense of additional computational cost associated with Hessian implementation.

Figures \ref{BP_data}a–d display seismograms computed using the initial model and the models obtained by the three inversion methods. Their differences with respect to the true seismograms are shown in Figures \ref{BP_data}e–h. These comparisons clearly demonstrate that, under identical inversion settings, implementing the Hessian produces velocity models that fit the data more accurately across all times and offsets. The improvement gained by implementing the Hessian, compared to the diagonal approximation, is clearly evident in Figures \ref{BP_data}g and \ref{BP_data}h. The Hessian-based inversion yields a model that more accurately reproduces the late arrivals and long-offset events.

\begin{figure}[!htb]
    \centering
    \includegraphics[width=.5\linewidth]{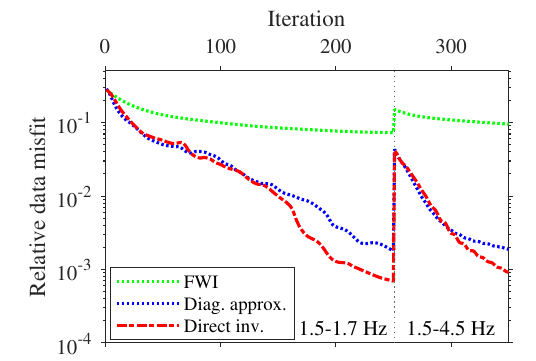}
    \caption{Relative data misfit trajectories for the 2004 BP salt inversion. Green: reduced FWI, blue: diagonal Hessian approximation, red: proposed direct Hessian inversion.}
    \label{BP_curve}
\end{figure}

\begin{figure}[!h]
    \centering
    \includegraphics[width=.4\linewidth]{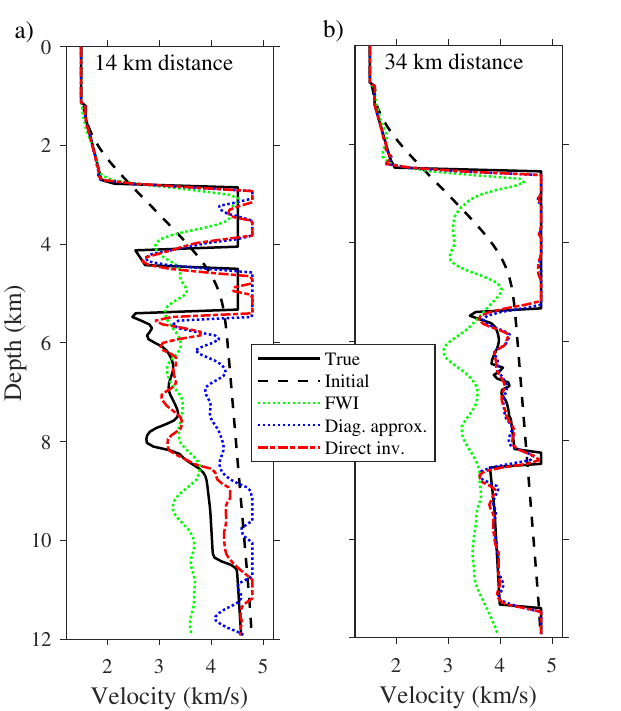}
    \caption{Vertical velocity logs extracted from \fref{BP_results}(a–e) at (a) 14 km and (b) 34 km distance.}
    \label{BP_log}
\end{figure}

\begin{figure}[!htb]
    \centering
    \includegraphics[width=1\linewidth]{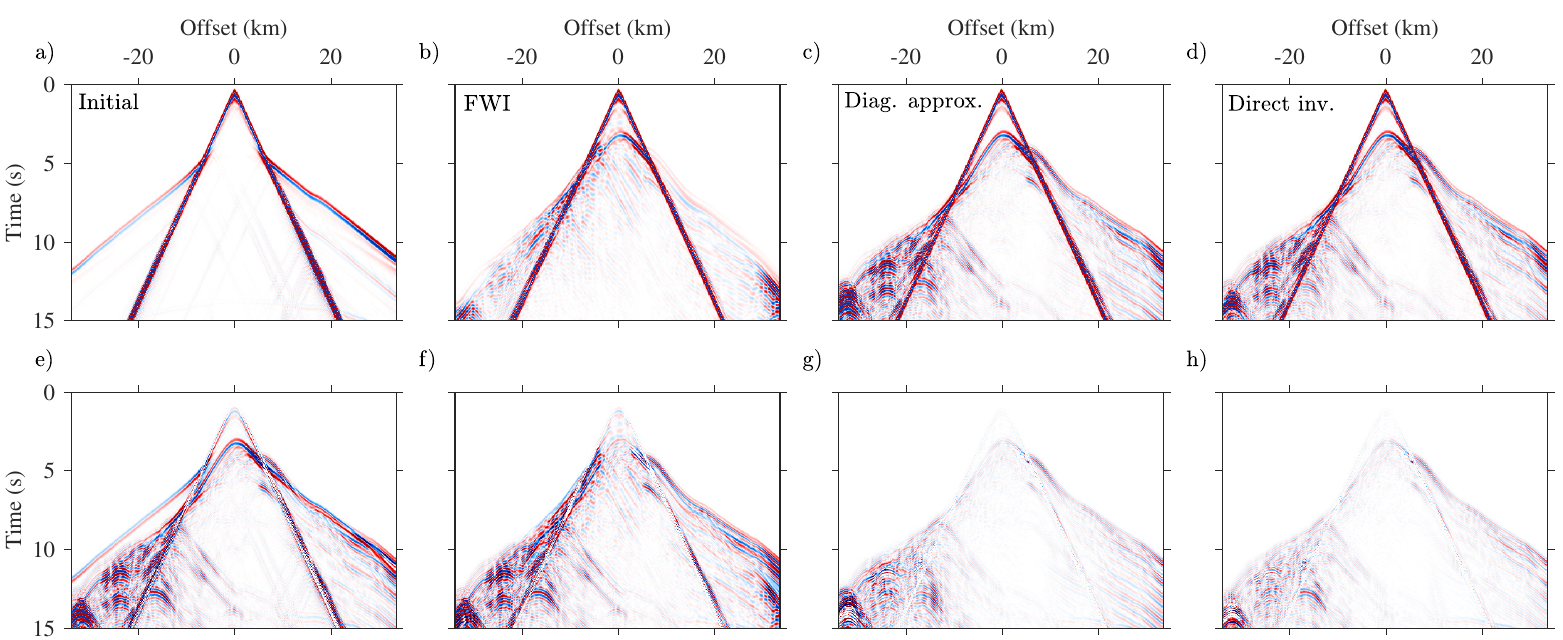}
    \caption{Seismograms computed for the 2004 BP salt model example using (a) the initial model and the inverted models obtained by (b) reduced FWI, (c) the diagonal Hessian approximation, and (d) the proposed direct Hessian inversion. (e–h) Residuals relative to true data.}
    \label{BP_data}
\end{figure}

Finally, to evaluate the performance of the proposed direct Hessian inversion against the exact Hessian, we conducted an inversion on a cropped portion of the 2004 BP salt model shown in Figure \ref{BP_crop_results}a, using the initial model in Figure \ref{BP_crop_results}b. We applied reduced FWI and multiplier-based methods using the diagonal Hessian approximation, the proposed direct Hessian inversion, and an approximation of the exact Hessian. Since computing the exact Hessian explicitly was infeasible, we used CG and GMRES iterative solvers to approximate the Hessian inverse at each iteration. We performed 100 CG iterations and 50 GMRES iterations.
Even for this relatively small test case, the computational cost of CG and GMRES was substantial, requiring 6000 and 3000 PDE solves per iteration, respectively, corresponding to CPU times of 2205 s and 1092 s per iteration. In contrast, the proposed direct Hessian inversion required only 100 PDE solves and 51 s per iteration (see Table \ref{Table_2}).

The inverted models obtained with the different approaches are shown in Figures \ref{BP_crop_results}c–g, with the corresponding convergence curves presented in Figure \ref{BP_crop_curve}. As before, reduced FWI became trapped in a local minimum, while all multiplier-based approaches successfully converged to geologically plausible results. The models obtained using the proposed direct Hessian inversion, CG, and GMRES are nearly indistinguishable visually, and this similarity is reflected in their convergence behavior. Figure \ref{BP_crop_curve} clearly demonstrates that incorporating the Hessian significantly accelerates convergence and improves reconstruction quality.
\begin{figure}[!htb]
    \centering
    \includegraphics[width=1\linewidth]{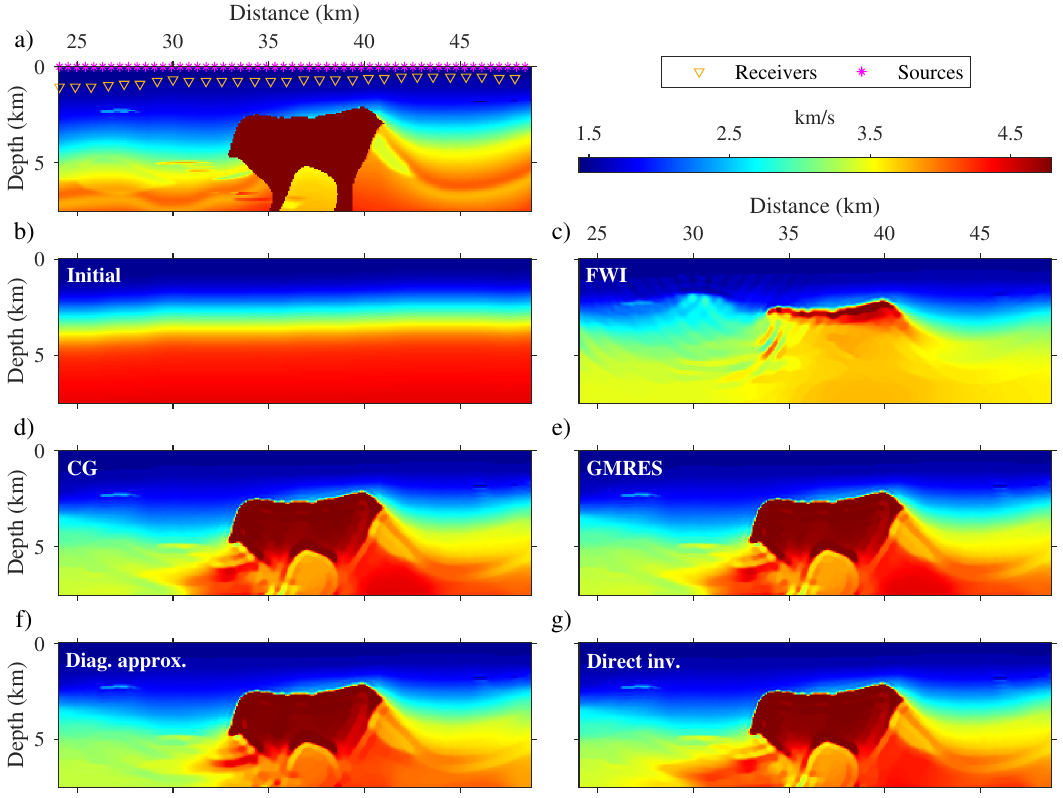}
    \caption{(a) Cropped BP model. (b) Initial model. (c) Reconstructed model using reduced FWI. (d–g) Reconstructed models using multiplier-based FWI with four different Hessian inversion strategies: (d) 100 CG iterations, (e) 50 GMRES iterations, (f) diagonal approximation, and (g) direct inversion.}
    \label{BP_crop_results}
\end{figure}

\begin{figure}[!htb]
    \centering
    \includegraphics[width=.5\linewidth]{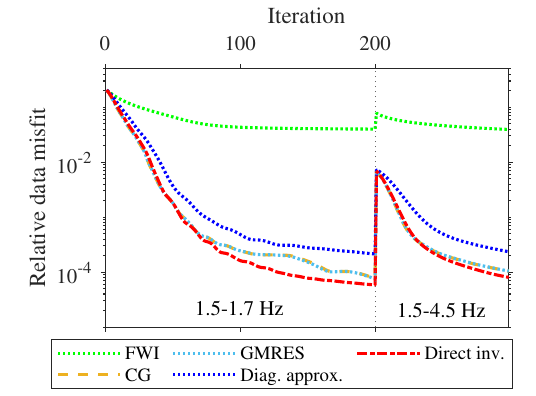}
    \caption{Relative data misfit trajectories for the cropped BP inversion by different methods (\fref{BP_crop_results}).}
    \label{BP_crop_curve}
\end{figure}

\section{DISCUSSION}
In this study, we develop a numerical algorithm for the direct inversion of the data-space Hessian arising in time-domain extended-space FWI. Specifically, we implement the inversion within the multiplier-based FWI framework, following the augmented Lagrangian formulation introduced by \citep{Gholami_2022_EFW}. However, the proposed inversion approach is general and can also be applied to other FWI formulations, including contrast-source inversion \citep{Abubakar_2009_FDC}, wavefield reconstruction inversion \citep{vanLeeuwen_2016_PMP}, matched-source inversion \citep{Huang_2016_MSW}, or even standard reduced FWI using the split Gauss–Newton Hessian \citep{Gholami_2024_FWI}.

In multiplier-based FWI, the data-space Hessian enables an accurate estimation of the data-side Lagrange multipliers, which serve both as adjoint sources for back-propagating wavefields and, once projected into the subsurface, as source extensions for the forward wavefield. Accurate Hessian inversion improves the quality of these multiplier estimates, potentially accelerating the convergence of the inversion algorithm.

The method presented here is based on a block-diagonal approximation of the Hessian in the Fourier domain. Numerical examples demonstrate that this approximation is sufficiently accurate in practice. To construct the frequency-domain block diagonals, we first compute the Green functions for all active receivers (following the \textit{Randomized receiver inversion} subsection) and then calculate the weighted correlation vectors $\bold{q}^{i,j}$ to assemble the frequency-domain Hessian blocks $\hat{\bold{Q}}(\omega)$ (see Algorithm \ref{alg:Hessian}).

For large-scale 3D problems, the computation of $\bold{q}^{i,j}$ remains a major bottleneck. A straightforward implementation would require storing all wavefields at a cost of $O(N_r N_t N_x)$ memory, which is typically prohibitive, while forming $\bold{q}^{i,j}$ scales as $O(N_x N_r^2 N_t^2)$. The use of randomized receiver inversion and storing Green functions at the Nyquist rate alleviates these demands to some extent, but memory and computational costs remain substantial.
Several strategies could be considered to mitigate these challenges. In principle, $\bold{q}^{i,j}$ could be computed on-the-fly during wave propagation, thus avoiding the need to store full wavefields, although the feasibility of this approach remains uncertain. Another possibility is to write each wavefield to disk during propagation and compute the cross-correlations in a post-processing stage by sequentially reading them back. This avoids keeping all wavefields in memory simultaneously but introduces significant I/O overhead and storage requirements, particularly in 3D. A further alternative is to employ lossless compression algorithms to reduce storage
\citep{Sun_2013_TWO,Zand_2019_CIR,Lee_2022_IME}. 
Let the $i$-th Green function $\bold{g}^i$ be reshaped into a matrix $\bold{G}^i$, where each column corresponds to a trace recorded at a spatial location:
\begin{equation}
    \bold{G}^i=
    \begin{bmatrix}
        \bold{g}^i(:, \bold{x}_1) & \bold{g}^i(:, \bold{x}_2) & \cdots & \bold{g}^i(:, \bold{x}_{N_x})
    \end{bmatrix}.
\end{equation}
The cross-correlation matrix can then be written as
\begin{equation}
    \bold{G}^{i,j}=\sum_{\bold{x}} \bold{g}^i(:, \bold{x}) \bold{g}^j(:, \bold{x})^T = \bold{G}^i(\bold{G}^j)^T.
\end{equation}
Now let $\boldsymbol{\Psi}$ be a suitable orthonormal transform (e.g., discrete cosine transform or a wavelet basis) that effectively compresses the spatial dimension of the Green functions. Since $\boldsymbol{\Psi}^T \boldsymbol{\Psi} = \bold{I}$, we can equivalently express
\begin{equation}
    \bold{G}^{i,j}=\bold{G}^i\bold{I}(\bold{G}^j)^T=\bold{G}^i\boldsymbol{\Psi}^T\boldsymbol{\Psi}(\bold{G}^j)^T=\widetilde{\bold{G}}^i(\widetilde{\bold{G}}^j)^T.
\end{equation}
where $\widetilde{\bold{G}}^i = \bold{G}^i \boldsymbol{\Psi}^T$ and $\widetilde{\bold{G}}^j = \bold{G}^j \boldsymbol{\Psi}^T$ are the transformed Green function matrices.
This transformation concentrates most of the spatial energy into a small number of coefficients, leaving many columns of $\widetilde{\bold{G}}^i$ and $\widetilde{\bold{G}}^j$ nearly zero. Thus, only the leading $\widetilde{N}_x \ll N_x$ columns need to be retained without significantly affecting the cross-correlation results. Consequently, the storage per time slice is reduced from $N_x$ to $\widetilde{N}_x$, and the computational cost decreases proportionally by a factor of $\widetilde{N}_x / N_x$.

Future work will also focus on extending the proposed direct Hessian inversion method to multiplier-based time-domain elastic FWI, where implementing the Hessian can be even more crucial for suppressing cross-talk noise, and to applications in 3D. To circumvent the prohibitive costs of full 3D Hessian computation, it may not be necessary to reconstruct the Hessian at every iteration; instead, updating it every few iterations could be sufficient to achieve reasonable convergence. Since the Lagrange multiplier estimates are inherently approximate, applying the Hessian inverse mainly improves their quality, which can accelerate convergence. One practical strategy is to implement the Hessian inversion in the early iterations and then switch to the diagonal approximation proposed in \citep{Gholami_2022_EFW}. Alternatively, the dual form of the algorithm could be employed \citep{Aghazade_2025_FAF}, where the model parameters remain fixed and the focus shifts to accurately constructing the multipliers. In this dual framework, because the background model does not change, the Hessian needs to be constructed only once.

\section{Conclusions}
We proposed an efficient hybrid-domain approach to significantly reduce the computational cost of time-domain extended-source FWI. The method constructs the data-space Hessian matrix in the time domain, leveraging time-domain wave equations solvers, and directly inverts it in the frequency domain by exploiting the frequency separability, allowing independent solutions for each frequency. This formulation allows the simultaneous use of the Hessian across multiple sources, leveraging the computational advantages inherent in both the time and frequency domains. Numerical experiments on the Marmousi II and 2004 BP salt models demonstrate the substantial computational gains and accuracy improvements achieved by the proposed method. These results highlight its effectiveness and practicality, making it a promising solution for large-scale extended-source FWI applications in complex geophysical models.

\section{ACKNOWLEDGMENTS}  
This research was partially funded by the SONATA BIS (grant no. 2022/46/E/ST10/00266) of the National Science Center in Poland.

\bibliographystyle{seg}

\begin{thebibliography}{}
\itemsep0pt

\bibitem[Abubakar et~al., 2009]{Abubakar_2009_FDC}
Abubakar, A., W. Hu, T.~M. Habashy, and P.~M. {van den Berg},  2009,
  Application of the finite-difference contrast-source inversion algorithm to
  seismic full-waveform data: Geophysics, {\bf 74}, WCC47--WCC58.

\bibitem[Aghamiry et~al., 2019]{Aghamiry_2019_IWR}
Aghamiry, H., A. Gholami, and S. Operto,  2019, Improving full-waveform
  inversion by wavefield reconstruction with alternating direction method of
  multipliers: Geophysics, {\bf 84(1)}, R139--R162.

\bibitem[Aghamiry et~al., 2020]{Aghamiry_2020_AED}
Aghamiry, H.~S., A. Gholami, and S. Operto,  2020, Accurate and efficient
  data-assimilated wavefield reconstruction in the time domain: Geophysics,
  {\bf 85}, A7--A12.

\bibitem[Aghazade et~al., 2021]{Aghazade_2021_RSE}
Aghazade, K., H.~S. Aghamiry, A. Gholami, and S. Operto,  2021, Randomized
  source sketching for full waveform inversion: IEEE Transactions on Geoscience
  and Remote Sensing, {\bf 60}, 1--12.

\bibitem[Aghazade and Gholami, 2025]{Aghazade_2025_FAF}
Aghazade, K. and A. Gholami,  2025, Fast and automatic full waveform inversion
  by dual augmented {L}agrangian: Computational Geosciences, {\bf 29}, 1--22.
  
\bibitem[Barnier et~al., 2023a]{Barnier_2023_FWIp}
Barnier, G., E. Biondi, R.~G. Clapp, and B. Biondi,  2023a, Full waveform
  inversion by model extension: practical applications: Geophysics, {\bf 88},
  1--138.

\bibitem[Barnier et~al., 2023b]{Barnier_2023_FWIt}
--------, 2023b, Full waveform inversion by model extension: theory, design and
  optimization: Geophysics, {\bf 88}, 1--206.
  
  \bibitem[Ben-Hadj-Ali  et~al., 2011]{Ben_2011_EFF}Ben-Hadj-Ali, H., Operto, S. \& Virieux, J. An efficient frequency-domain full waveform inversion method using simultaneous encoded sources. {\em Geophysics}. \textbf{76}, R109-R124 (2011)

\bibitem[Boonyasiriwat et~al., 2009]{Boonyasiriwat_2009_EMM}
Boonyasiriwat, C., P. Valasek, P. Routh, W. Cao, G.~T. Schuster, and B. Macy,
  2009, An efficient multiscale method for time-domain waveform tomography:
  Geophysics, {\bf 74}, WCC59--WCC68.

\bibitem[Bunks et~al., 1995]{Bunks_1995_MSW}
Bunks, C., F.~M. Salek, S. Zaleski, and G. Chavent,  1995, Multiscale seismic
  waveform inversion: Geophysics, {\bf 60}, 1457--1473.

\bibitem[Castellanos et~al., 2015]{Castellanos_2015_FFW1}
Castellanos, C., L. M\'etivier, S. Operto, R. Brossier, and J. Virieux,  2015,
  Fast full waveform inversion with source encoding and second-order
  optimization methods: Geophysical Journal International, {\bf 200(2)},
  720--744.

\bibitem[Ernst and Gander, 2011]{Ernst_2011_WDS}
Ernst, O.~G. and M.~J. Gander,  2011, Why it is difficult to solve {H}elmholtz
  problems with classical iterative methods: Numerical analysis of multiscale
  problems,  325--363.

\bibitem[Gholami et~al., 2022]{Gholami_2022_EFW}
Gholami, A., H.~S. Aghamiry, and S. Operto,  2022, Extended-space full-waveform
  inversion in the time domain with the augmented lagrangian method:
  Geophysics, {\bf 87}, R63--R77.

\bibitem[Gholami et~al., 2023]{Gholami_2023_MWI}
--------, 2023, Multiplier waveform inversion ({MWI}): A reduced-space {FWI} by
  the method of multipliers: Geophysics, {\bf 88}, R339–R354.

\bibitem[Gholami and Aghazade, 2024]{Gholami_2024_FWI}
Gholami, A. and K. Aghazade,  2024, Full waveform inversion and lagrange
  multipliers: Geophysical Journal International, {\bf 238}, 109--131.

\bibitem[Gray, 2006]{Gray_2006_TCM}
Gray, R. M., 2006, Toeplitz and circulant matrices: A review: Foundations and Trends® in Communications and Information Theory, {\bf 2}, 155-239.

\bibitem[Guo et~al., 2024]{Guo_2024_TDE}
Guo, G., S. Operto, A. Gholami, and H.~S. Aghamiry,  2024, Time-domain
  extended-source full-waveform inversion: Algorithm and practical workflow:
  Geophysics, {\bf 89}, R73--R94.

\bibitem[Hestenes, 1969]{Hestenes_1969_MAG}
Hestenes, M.~R.,  1969, Multiplier and gradient methods: Journal of
  optimization theory and applications, {\bf 4}, 303--320.

\bibitem[Huang et~al., 2018]{Huang_2018_VSE}
Huang, G., R. Nammour, and W.~W. Symes,  2018, Volume source-based extended
  waveform inversion: Geophysics, {\bf 83}, R369--387.

\bibitem[Huang et~al., 2016]{Huang_2016_MSW}
Huang, G., W. Symes, and R. Nammour,  2016, Matched source waveform inversion:
  Space-time extension: SEG Technical Program Expanded Abstracts, 1426--1431.

\bibitem[Lailly, 1983]{Lailly_1983_SIP}
Lailly, P.,  1983, The seismic inverse problem as a sequence of before stack
  migrations: Conference on {I}nverse {S}cattering, Theory and application,
  Society for Industrial and Applied Mathematics, Philadelphia, Expanded
  Abstracts, 206--220.

\bibitem[Lee and Chung, 2022]{Lee_2022_IME}
Lee, D. and W. Chung,  2022, Improving the memory efficiency of rtm using both
  nyquist sampling and dct based on gpu: Journal of Geophysics and Engineering,
  {\bf 19}, 706--723.
  
\bibitem[Lin et~al., 2023]{Lin_2023_FWR}
Lin, Y., T. van Leeuwen, H. Liu, J. Sun, and L. Xing,  2023, A fast wavefield
  reconstruction inversion solution in the frequency domain: Geophysics, {\bf
  88}, R257–R267.

\bibitem[M{\'e}tivier and Brossier, 2022]{Metivier_2022_RES}
M{\'e}tivier, L. and R. Brossier,  2022, Receiver-extension strategy for
  time-domain full-waveform inversion using a relocalization approach:
  Geophysics, {\bf 87}, R13--R33.

\bibitem[M{\'e}tivier et~al., 2017]{Metivier_2017_TRU}
M{\'e}tivier, L., R. Brossier, S. Operto, and V. J.,  2017, Full waveform
  inversion and the truncated {N}ewton method: SIAM Review, {\bf 59}, 153--195.

\bibitem[Nocedal and Wright, 2006]{Nocedal_2006_NO}
Nocedal, J. and S.~J. Wright,  2006, Numerical optimization: Springer,
  2\textsuperscript{nd} edition.

\bibitem[Operto et~al., 2023]{Operto_2023_FWI}
Operto, S., A. Gholami, H.~S. Aghamiry, G. Guo, S. Beller, K.
  Aghazade, F. Mamfoumbi, L. Combe, and A. Ribodetti,  2023, Extending the
  search space of full-waveform inversion beyond the single-scattering born
  approximation: A tutorial review: Geophysics, {\bf 88}, 1--32.

\bibitem[Powell, 1969]{Powell_1969_NLC}
Powell, M.~J.,  1969, A method for nonlinear constraints in minimization
  problems: Optimization,  283--298.

\bibitem[Pratt et~al., 1998]{Pratt_1998_GNF}
Pratt, R.~G., C. Shin, and G.~J. Hicks,  1998, {G}auss-{N}ewton and full
  {N}ewton methods in frequency-space seismic waveform inversion: Geophysical
  Journal International, {\bf 133}, 341--362.


\bibitem[Rizzuti et~al., 2021]{Rizzuti_2021_ADF}
Rizzuti, G., M. Louboutin, R. Wang, and F.~J. Herrmann,  2021, A dual
  formulation of wavefield reconstruction inversion for large-scale seismic
  inversion: Geophysics, {\bf 86}, R879--R893.

\bibitem[Sirgue et~al., 2008]{Sirgue_2008_FDW}
Sirgue, L., J.~T. Etgen, and U. Albertin,  2008, 3{D} frequency domain waveform
  inversion using time domain finite difference methods: 70\textsuperscript{th}
  {EAGE}, Conference and Exhibition, Roma, {Italy}, Expanded Abstracts, F022.

\bibitem[Sonbolestan et~al., 2022]{Sonbolestan_2022_ORD}
Sonbolestan, M., H. Aghamiry, A. Gholami, and S. Operto,  2022, On the role of
  data-space hessian in wavefield reconstruction inversion: 83rd EAGE Annual
  Conference \& Exhibition, 1--5.

\bibitem[Sun and Fu, 2013]{Sun_2013_TWO}
Sun, W. and L.-Y. Fu,  2013, Two effective approaches to reduce data storage in
  reverse time migration: Computers \& Geosciences, {\bf 56}, 69--75.

\bibitem[Symes, 2020]{Symes_2020_WRI}
Symes, W.~W.,  2020, Wavefield reconstruction inversion: an example: Inverse
  Problems, {\bf 36}, 105010.

\bibitem[Symes et~al., 2020]{Symes_2020_FWI}
Symes, W.~W., H. Chen, and S.~E. Minkoff,  2020, Full-waveform inversion by
  source extension: Why it works: Presented at the SEG International Exposition
  and Annual Meeting.

\bibitem[Symes et~al., 2024]{Symes_2024_NIM}
--------, 2024, A numerical investigation of matched source waveform inversion
  applied to acoustic transmission data: arXiv preprint arXiv:2412.06074.

\bibitem[Tarantola, 1984]{Tarantola_1984_ISR}
Tarantola, A.,  1984, Inversion of seismic reflection data in the acoustic
  approximation: Geophysics, {\bf 49}, 1259--1266.

\bibitem[{van Leeuwen} and Herrmann, 2016]{vanLeeuwen_2016_PMP}
{van Leeuwen}, T. and F. Herrmann,  2016, A penalty method for
  {PDE}-constrained optimization in inverse problems: Inverse Problems, {\bf
  32(1)}, 1--26.

\bibitem[{van Leeuwen} and Herrmann, 2013]{VanLeeuwen_2013_MLM}
{van Leeuwen}, T. and F.~J. Herrmann,  2013, Mitigating local minima in
  full-waveform inversion by expanding the search space: Geophysical Journal
  International, {\bf 195(1)}, 661--667.

\bibitem[Warner and Guasch, 2016]{Warner_2016_AWI}
Warner, M. and L. Guasch,  2016, Adaptive waveform inversion: Theory:
  Geophysics, {\bf 81}, R429--R445.

\bibitem[Zand et~al., 2019]{Zand_2019_CIR}
Zand, T., A. Malcolm, A. Gholami, and A. Richardson,  2019, Compressed imaging
  to reduce storage in adjoint-state calculations: IEEE Transactions on
  Geoscience and Remote Sensing, {\bf 57}, 9236--9241.

\end{thebibliography}
\newcommand{\SortNoop}[1]{}

\end{document}